\pdfoutput=1
\documentclass[11pt,twoside,a4paper,cmspaper,final,collab]{cms-tdr}
\def\svnVersion{12b005b}\def\svnDate{2022/05/12}\def\cmsCernNoTag{CERN-EP-2021-228}\def\cmsCernDate{\today}\def\cmsMessage{Published in the Journal of High Energy Physics as \href{http://dx.doi.org/10.1007/JHEP04(2022)047}{\doi{10.1007/JHEP04(2022)047}.}}
\begin{document}\cmsNoteHeader{EXO-20-002}

\newcommand{\DYJ}{\ensuremath{\text{DY+jets}}\xspace}
\newcommand{\N}{\ensuremath{\mathrm{N}}\xspace}
\newcommand{\jet}{\ensuremath{\mathrm{j}}\xspace}
\newcommand{\Jet}{\ensuremath{\mathrm{J}}\xspace}
\newcommand{\mWR}{\ensuremath{m_{\PWR}}\xspace}
\newcommand{\mN}{\ensuremath{m_{\N}}\xspace}
\newcommand{\gR}{\ensuremath{g_{\mathrm{R}}}\xspace}
\newcommand{\gL}{\ensuremath{g_{\mathrm{L}}}\xspace}
\newcommand{\LSFiii}{\ensuremath{\mathrm{LSF_{3}}}\xspace}
\newcommand{\mSD}{\ensuremath{m_{\mathrm{SD}}}\xspace}
\newcommand{\mlljj}{\ensuremath{m_{\ell\ell\jet\jet}}\xspace}
\newcommand{\mlJ}{\ensuremath{m_{\ell\Jet}}\xspace}
\newcommand{\mll}{\ensuremath{m_{\ell\ell}}\xspace}
\newcommand{\tW}{\ensuremath{ \PQt\PW}\xspace}

\newcommand{\cmsTable}[1]{\resizebox{\textwidth}{!}{#1}}

\cmsNoteHeader{EXO-20-002} 
\title{Search for a right-handed \texorpdfstring{\PW}{W} boson and a heavy neutrino in proton-proton collisions at \texorpdfstring{$\sqrt{s} = 13\TeV$}{sqrt(s) = 13 TeV}} 

\author{The CMS Collaboration}

\date{\today}

\abstract{
A search is presented for a right-handed \PW boson (\PWR) and a heavy neutrino (\N), in a final state consisting of two same-flavor leptons ($\Pe\Pe$ or $\PGm\PGm$) and two quarks. The search is performed with the CMS experiment at the CERN LHC using a data sample of proton-proton collisions at a center-of-mass energy of 13\TeV corresponding to an integrated luminosity of $138\fbinv$. The search covers two regions of phase space, one where the decay products of the heavy neutrino are merged into a single large-area jet, and one where the decay products are well separated. The expected signal is characterized by an excess in the invariant mass distribution of the final-state objects. No significant excess over the standard model background expectations is observed. The observations are interpreted as upper limits on the product of \PWR production cross sections  and branching fractions assuming that couplings are identical to those of the standard model \PW boson. For \N masses \mN equal to half the \PWR mass \mWR ($\mN=0.2\TeV$), \mWR is excluded at 95\% confidence level up to 4.7 (4.8) and 5.0 (5.4)\TeV for the electron and muon channels, respectively. This analysis provides the most stringent limits on the \PWR mass to date. 
}

\hypersetup{
pdfauthor={CMS Collaboration},
pdftitle={Search for a right-handed W boson and a heavy neutrino in proton-proton collisions at sqrt(s)=13 TeV},
pdfsubject={CMS},
pdfkeywords={CMS, exotica}}

\maketitle

\section{Introduction}
\label{sec:intro}

Left-right (LR) symmetric models~\cite{lr, lr1, lr2, lr3}, which extend the electroweak sector of the standard model (SM) by a right-handed $\mathrm{SU}(2)$ group, provide a possible explanation for parity violation in the SM as the consequence of spontaneous symmetry breaking at a multi-\TeVns mass scale. 
These models predict a heavy partner of the SM \PW boson, a heavy, right-handed gauge boson \PWR that is coupled to right-handed fermions. 
In addition, LR models also provide an explanation for the small mass of SM neutrinos through the seesaw mechanism~\cite{PhysRevLett.44.912,Das:2017hmg,GellMann:1980vs}, which requires the existence of a heavy right-handed neutrino (\N) for each lepton flavor. 
The heavy neutrinos couple exclusively to leptons of the corresponding flavor, and have different masses.

The coupling strength of the \PWR boson to the SM particles (\gR) is a free parameter in most LR models. 
We assume LR symmetry, such that \gR is the same as the SM coupling constant \gL.

\begin{figure}[ht]
\centering
  \includegraphics[width=0.70\textwidth]{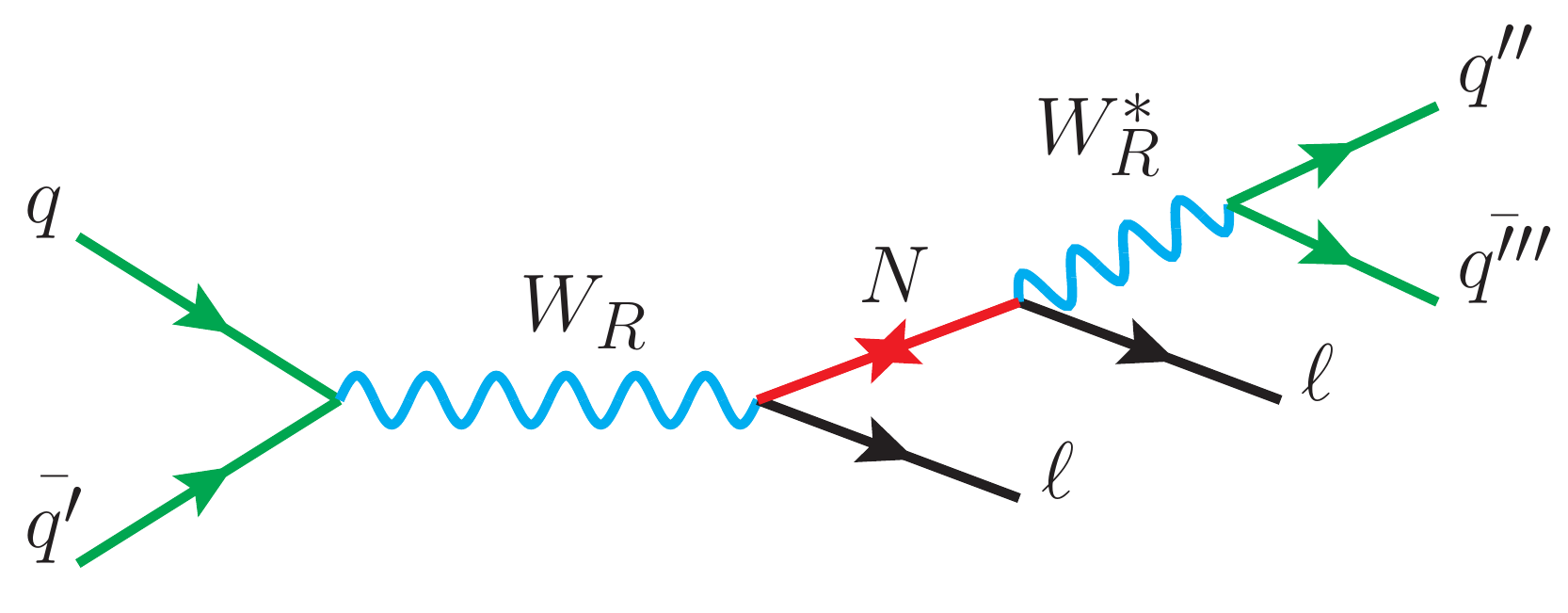}
  \caption{Feynman diagram for the production of a heavy neutrino via the decay of a \PWR boson.}
\label{fig:feynman}
\end{figure}

The dominant production process for the \PWR boson at the CERN LHC is the Drell--Yan (DY) mechanism.
The leading order Feynman diagram for this process is shown in Fig.~\ref{fig:feynman}.
Although the potential Majorana nature of the right-handed neutrinos implies that the final-state charged leptons can have the same sign, we do not impose charge requirements on the final-state leptons, in order to remain sensitive to the widest possible range of models.
Searches for \PWR bosons and heavy neutrinos have been performed by the ATLAS~\cite{Aad:2015xaa, Aaboud:2018spl, Aaboud:2019wfg} and CMS~\cite{Khachatryan:2014dka, Sirunyan:2018pom, Sirunyan:2018xiv, Khachatryan:2016jqo, Sirunyan:2017yrk} Collaborations using LHC proton-proton ($\Pp\Pp$) collision data at $\sqrt{s}=8$ and 13\TeV. 
These searches have excluded regions of phase space with masses of the right-handed \PW boson and heavy neutrino (\mWR and \mN, respectively) up to several \TeVns. 

In this paper, we extend the region of parameter space covered in the previous 2016 CMS search for \PWR bosons in events with two same-flavor leptons (\Pe or \PGm) and two jets using 2016 data~\cite{Sirunyan:2018pom}, by including the regime where the \PWR boson is heavy compared to the \N ($\mWR/\mN\geq 10$).
In this scenario, the heavy neutrinos are produced with large transverse momentum (\pt) and their decay products are collimated along their direction of motion. 
Therefore, the heavy neutrino decay is reconstructed in a single jet and identified using jet substructure techniques, similar to those discussed in Refs.~\cite{PhysRevD.62.013001, Mattelaer:2016ynf, PhysRevD.94.095016} and previously applied in Ref.~\cite{Aaboud:2019wfg}; we refer to these as ``boosted'' events, in contrast to ``resolved'' events where the two jets from the heavy neutrino decay are reconstructed separately.
The inclusion of boosted events leads to significant improvements in the search sensitivity in the region where $\mN < 0.5\TeV$. 
To obtain maximum sensitivity, a statistical combination of the resolved and boosted results is performed. 
For this analysis we use the data sample collected in 2016--2018 with the CMS detector at the LHC corresponding to an integrated luminosity of $138\fbinv$.
Tabulated results are provided in HEPData~\cite{hepdata}.

\section{The CMS detector}
\label{sec:CMSdetector}

A detailed description of the CMS detector, together with a definition of the coordinate system and the relevant kinematic variables, can be found in Ref.~\cite{Chatrchyan:2008zzk}. The central feature of the CMS apparatus is a superconducting solenoid of 6\unit{m} internal diameter, providing a magnetic field of 3.8\unit{T}. Within the solenoid volume are a silicon pixel and strip tracker, a lead tungstate crystal electromagnetic calorimeter (ECAL), and a brass and scintillator hadron calorimeter (HCAL), each composed of a barrel and two endcap sections. Forward calorimeters extend the pseudorapidity ($\eta$) coverage provided by the barrel and endcap detectors. Muons are detected in gas-ionization chambers embedded in the steel flux-return yoke outside the solenoid.

Events of interest are selected using a two-tiered trigger system~\cite{Khachatryan:2016bia}. The first level, composed of custom hardware processors, uses information from the calorimeters and muon detectors to select events at a rate of around 100\unit{kHz} within a fixed time interval of less than $4\mus$~\cite{Sirunyan:2020zal}.
The second level, known as the high-level trigger, consists of a farm of processors running a version of the full event reconstruction software optimized for fast processing, and reduces the event rate to around 1\unit{kHz} before data storage.

\section{Simulated samples}
\label{sec:simSamples}

All signal events are simulated using the \MGvATNLO 2.6.5~\cite{MadGraph} Monte Carlo (MC) event generator 
at leading order (LO) following the prescriptions described in Refs.~\cite{Mattelaer:2016ynf,PhysRevD.94.095016} 
for various \mWR hypotheses in the range from 0.2 to 7.0\TeV, with \mN ranging from 0.1\TeV up to \mWR where \N is a Majorana neutrino. 
The events are generated in final states with two same-flavor (SF) leptons ($\Pe\Pe$ or $\PGm\PGm$). 
The production cross sections are scaled to next-to-LO (NLO) in quantum chromodynamics (QCD) using $K$ factors obtained from the same generator.

The background samples are simulated with several MC event generators.
The \PZ and \PW boson production associated with jets is simulated with the \MGvATNLO 2.2.2 (2.4.2) generator for 
up to four parton-level jets at LO for the 2016 (2017--2018) samples,
with the MLM matching scheme between jets from matrix element calculations and parton showers~\cite{Alwall:2007fs}. 
The \POWHEG 2.0~\cite{Frixione:2007vw, Re:2010bp, Alioli:2009je, ttbar2007} generator is used to model the \ttbar production, as well as \tW and $t$-channel single top quark production at NLO.
The $s$-channel single top quark production is generated with the \MGvATNLO generator at NLO precision.
The inclusive decay of \ttbar produced in association with a \PW and \PZ boson is simulated by \MGvATNLO at NLO and LO precision, respectively.
The \PYTHIA generator~\cite{Sjostrand:2014zea} is used to simulate diboson processes ($\PW\PW$, $\PW\PZ$, and $\PZ\PZ$) at LO.
Triple vector boson ($\PW\PW\PW$, $\PW\PW\PZ$, $\PW\PZ\PZ$, and $\PZ\PZ\PZ$) events are generated at NLO using \MGvATNLO.

The generators used for the signal and background processes are interfaced with \PYTHIA 8.226 (8.230)~\cite{Sjostrand:2014zea} to simulate the parton showering and hadronization in the 2016 (2017--2018) samples. 
The \PYTHIA parameters for the underlying event description are set with either the CUETP8M1~\cite{Khachatryan:2015pea} tune for the 2016 samples or the CP5~\cite{Sirunyan:2019dfx} tune for the 2017--2018 samples.
For the description of the parton distribution functions (PDFs), 
the NNPDF3.0~\cite{Ball:2014uwa} PDF sets are used to produce all simulated background samples in 2016. 
For the 2017--2018 background samples, the NNPDF3.1 next-to-NLO (NNLO) PDF sets~\cite{Ball:2017nwa} are used. 
The NNPDF3.1 NNLO PDF sets are used to simulate all the signal samples.

All simulated samples were processed through a \GEANTfour simulation~\cite{Agostinelli:2002hh} of the CMS detector.
Multiple $\Pp\Pp$ collisions may occur in the same or adjacent LHC bunch crossings (pileup) and contribute to the overall event activity in the detector. 
This effect is included in the simulation, and the distribution of the number of pileup interactions is adjusted to match the one observed in the data, assuming a total inelastic cross section of 69.2\unit{mb}~\cite{Sirunyan:2018nqx}.

\section{Event reconstruction and object selection}
\label{sec:EventSelection}

For this search, the events were selected at the trigger level by requiring the presence of a high-momentum lepton in the event. 
A combination of three triggers that require 
an isolated electron with $\pt>27\,(32)\GeV$,
an electron with $\pt>115\GeV$,
or a photon with $\pt>175\,(200)\GeV$
is used to collect events that contain at least one electron in 2016 (2018).
In 2017, two triggers with the requirement of an isolated electron with $\pt>35\GeV$ or a photon with $\pt>200\GeV$ were used.
The single-electron triggers differ in their usage of isolation requirements: while the lower-threshold trigger requires electrons to be well isolated, the higher-threshold trigger does not, which gives an improved efficiency at high \pt.
Similarly, the single-photon trigger avoids reliance on the online track reconstruction and increases the overall efficiency for electrons with $\pt>200\GeV$.
Events containing at least one isolated muon are collected by triggers that require a muon with a minimum \pt of 50\GeV.

The candidate vertex with the largest value of summed physics-object $\pt^2$ is taken to be the primary $\Pp\Pp$ interaction vertex.
The physics objects used for this determination are the jets, clustered using the jet finding algorithm~\cite{Cacciari:2008gp,Cacciari:2011ma} with the tracks assigned to candidate vertices as inputs, as well as the remaining single tracks (including identified leptons), and the associated missing transverse momentum, taken as the negative vector sum of the \pt of those objects.

The global event reconstruction (also called particle-flow event reconstruction)~\cite{CMS-PRF-14-001} aims to reconstruct and identify each individual particle in an event, 
with an optimized combination of all subdetector information.
In this process, the identification of the particle type (photon, electron, muon, charged and neutral hadron) plays an important role in the determination of the particle direction and energy. Photons (\eg, coming from \Pgpz decays or from electron bremsstrahlung) are identified as ECAL energy clusters not linked to the extrapolation of any charged particle trajectory to the ECAL. Each electron (\eg, coming from photon conversions in the tracker material or from \PB hadron semileptonic decays) is identified as a primary charged particle track and potentially many ECAL energy clusters corresponding to this track extrapolation to the ECAL and to possible bremsstrahlung photons emitted along the way through the tracker material. Muons (\eg, from \PB hadron semileptonic decays) are identified as tracks in the central tracker consistent with either a track or several hits in the muon system, and associated with calorimeter deposits compatible with the muon hypothesis. Charged hadrons are identified as charged particle tracks neither identified as electrons, nor as muons. Finally, neutral hadrons are identified as HCAL energy clusters not linked to any charged hadron trajectory, or as a combined ECAL and HCAL energy excess with respect to the expected charged hadron energy deposit. 

The energy of each photon is obtained from the ECAL measurement. 
The energy of each electron is determined from a combination of the electron momentum at the primary interaction vertex as determined by the tracker, 
the energy of the corresponding ECAL cluster, and the energy sum of all bremsstrahlung photons spatially compatible with originating from the electron track. 
The momentum of each muon is obtained from the trajectory of the corresponding track
in the magnetic field. The energy of each charged hadron is determined from a combination of
the momentum measured in the tracker and the matching ECAL and HCAL energy deposits,
corrected for the response function of the calorimeters to hadronic showers. Finally, the energy
of each neutral hadron is obtained from the corresponding corrected ECAL and HCAL energies.

For each event, the jets are clustered from the reconstructed particles using the infrared and collinear safe anti-\kt algorithm~\cite{Cacciari:2008gp, Cacciari:2011ma} with a distance parameter of 0.4 (AK4 jets) or 0.8 (AK8 jets). 
The jet momentum is determined as the vectorial sum of all particle momenta in the jet, and is found from simulation to be, on average, within 5--10\% of the true momentum over the whole \pt spectrum and detector acceptance~\cite{Khachatryan:2016kdb}. 

Pileup interactions can contribute additional tracks and calorimetric energy deposits to the reconstructed jets, increasing their apparent momentum. 
To mitigate this effect for AK4 jets, charged particles identified as originating from pileup vertices are discarded and an offset correction is applied to correct for the remaining contributions. 
Jet energy corrections are derived by comparing the average measured energies of jets in data with those of simulated jets.
In situ measurements of the momentum balance in dijet, $\text{photon}\text{+jet}$, $\PZ\text{+}\text{jet}$, and multijet events are used to account for any residual differences in jet energy scale in data and simulation~\cite{Khachatryan:2016kdb}. The typical jet energy resolution is 15\% at 10\GeV, 8\% at 100\GeV, and 4\% at 1\TeV. Additional selection criteria are applied to each jet to remove jets potentially dominated by anomalous contributions from various subdetector components or reconstruction failures~\cite{CMS-PAS-JME-16-003}.
For AK8 jets, the pileup-per-particle identification algorithm~\cite{Bertolini:2014bba,Sirunyan:2020foa} is used to mitigate the effect of pileup at the reconstructed-particle level, making use of local shape information, event pileup properties and tracking information. 

The momentum resolution for electrons with $\pt \approx 45\GeV$ from $\PZ \to \Pe \Pe$ decays ranges from 1.7\% to 4.5\%. It is generally better in the barrel region than in the endcaps, and also depends on the bremsstrahlung energy emitted by the electron as it traverses the material in front of the ECAL~\cite{Khachatryan:2015hwa}.

Muons are measured in the range $\abs{\eta} < 2.4$, with detection planes made using three technologies: drift tubes, cathode strip chambers, and resistive-plate chambers. The single-muon trigger efficiency exceeds 90\% over the full $\eta$ range, and the subsequent efficiency to reconstruct and identify muons is greater than 96\%. 
Matching muons to tracks measured in the silicon tracker results in a fractional \pt resolution of 1\% in the barrel and 3\% in the endcaps for muons with \pt up to 100\GeV, and of better than 7\% in the barrel for muons with \pt up to 1\TeV~\cite{Sirunyan:2018fpa}.

To reconstruct resolved \PWR candidates, events with two leptons and at least two AK4 jets are selected.
Events with additional leptons are rejected, and if more than two jets exist, the two jets with the highest \pt are used.
The leading (subleading) lepton is required to have $\pt > 60 (53)\GeV$ and to be within the fiducial acceptance ($\abs{\eta} < 2.4$). 
Electrons are rejected if the cluster lies in the range $1.44 < \abs{\eta} < 1.57$, which corresponds to the transition region between the barrel and endcap sections of the ECAL, where the performance is degraded. 
To suppress muons originating from hadron decays or pion punch-through in jets, 
the \pt sum of additional tracks that originate at the PV and are inside a cone of $\Delta R<0.3$, where $\Delta R\equiv\sqrt{\smash[b]{(\Delta\eta)^2 + (\Delta\phi)^2}}$, is required to be less than 10\% of the muon \pt.
Electrons are also required to be isolated, \ie, the \pt sum of all tracks in a cone of $\Delta R<0.3$ centered on the electron candidate, not associated with the electron and originating from the PV, must be below 5\GeV.
Dedicated identification algorithms, optimized for the selection of high-momentum leptons~\cite{Khachatryan:2015hwa, Chatrchyan:2012xi} are used. 
The two jet candidates must each have $\pt > 40\GeV$ and be within $\abs{\eta} < 2.4$. 
To avoid overlaps between leptons and jets, they are required to be separated in $\Delta R$ by at least 0.4.
An event is considered resolved if it contains four well-separated final-state objects as described above.

Events that fail the resolved selection criteria are used to reconstruct the boosted \PWR candidates.
The $\N\to \ell \cPq\overline{\cPq'}$ decay is reconstructed as a single AK8 jet.
The boosted signature requires an AK8 jet with $\pt > 200\GeV$ and two leptons that have the same \pt requirements as in the resolved search. 
The leading lepton is required to be separated in the azimuthal direction by $\Delta \phi > 2.0$ with respect to the AK8 jet.
The subleading lepton, with the isolation requirement removed, is required to fall within the AK8 jet cone. 
The AK8 jet with the highest \pt is chosen when multiple jets satisfy the conditions above, and events with additional isolated leptons are rejected.
The jet is groomed~\cite{Salam:2009jx} to remove soft and wide-angle radiation using the soft-drop algorithm~\cite{Dasgupta:2013ihk,Larkoski:2014wba}, with the soft radiation fraction parameter $z$ set to 0.1 and the angular exponent parameter $\beta$ set to 0.
The groomed jet is used to compute the soft-drop jet mass (\mSD) and each AK8 jet is required to have $\mSD > 40\GeV$, to optimize the signal selection and background rejection.

The lepton subjet fraction (\LSFiii) algorithm~\cite{Brust:2014gia} is used to determine the consistency of the jet with three subjets, where one subjet is dominated by the four-momentum of the lepton.
This algorithm clusters the constituents of a jet into three subjets using the exclusive-\kt algorithm, and \LSFiii is defined as the ratio of the lepton \pt to its associated subjet \pt ($\pt(\ell)/\pt(\text{subjet})$).
Jets with higher \LSFiii values are considered to have a more isolated lepton within the jet and therefore the lepton is assumed to originate from a prompt decay.
A selection of $\LSFiii > 0.75$ is required for all signal AK8 jets,
which removes more than 81 (94)\% of dielectron (dimuon) background events that have leptons originating from the decay of particles with non-negligible lifetimes (nonprompt leptons),
while keeping 87 (92)\% of signal events with $(\mWR,\mN)=(5.0,0.2)\TeV$.

For the resolved events we reconstruct an invariant mass distribution from the two leptons and two AK4 jets (\mlljj), and for the boosted events we consider the distribution in invariant mass of the lead lepton and AK8 jet (\mlJ). 
We search for deviations from the expected SM background in these distributions in the mass regions with \mlljj or $\mlJ > 0.8\TeV$.
To reduce the contribution from \PZ boson production, we also impose a requirement of $\mll > 0.4\,(0.2)\TeV$ in the resolved (boosted) search.
For the signal samples with $\mWR=5.0\TeV$ and $\mN=0.2\,(3.0)\TeV$, 
the product of acceptance and efficiency is 21--22 (49--50)\% and 34--43 (54--67)\% in the dielectron and dimuon boosted (resolved) signal regions (SRs), respectively,
depending on the year.

\section{Background estimation}
\label{sec:bkgdEst}

\begin{figure}[htbp]
  \centering
  \includegraphics[width=0.8\textwidth]{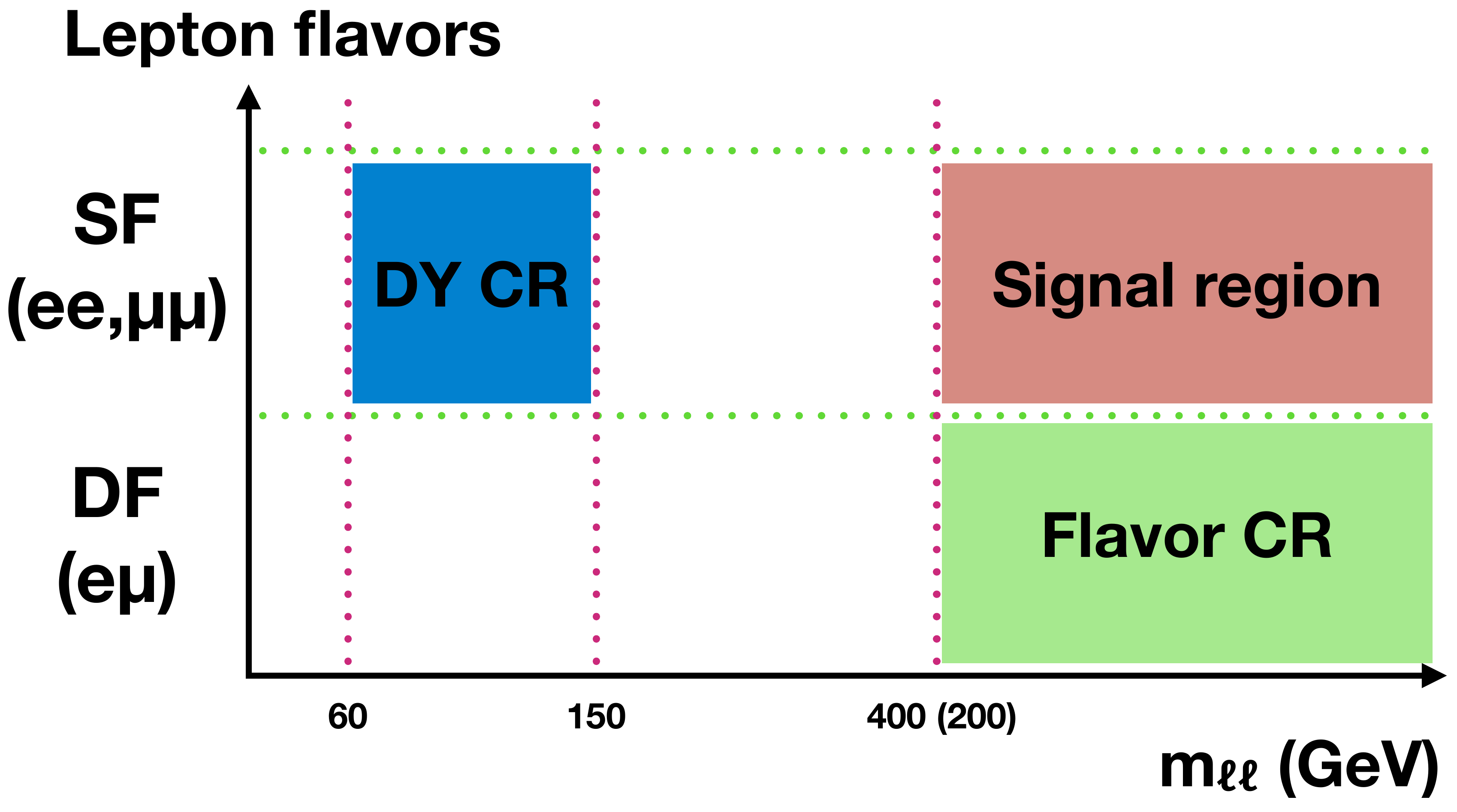}
  \caption{
    A schematic diagram of the analysis region. 
    The minimum values of the dilepton mass in the SR and the flavor CR is 400 (200)\GeV for the resolved (boosted) region.
    The backgrounds from \tW and \ttbar production are estimated from the flavor CR (green), where different flavor (DF) leptons are required.
    The DY background is estimated from the DY CR (blue).
  }
  \label{fig:BkgdSchem}
\end{figure}

The dominant SM processes that contribute to the background in each SR are 
DY production of lepton pairs with additional jets in the final state (\DYJ), 
leptonic decays of pair-produced top quarks, 
and single top production with an associated \PW boson.
These backgrounds are estimated from simulation and the modeling is corrected using control regions (CRs), as described below.
A schematic diagram presenting the CRs and SR is shown in Fig.~\ref{fig:BkgdSchem}.

The \ttbar events with at least one hadronically decaying \PW boson, 
$s$- and $t$-channel production of a single top quark,
and \PW boson production in association with jets can contribute to the SRs 
when nonprompt leptons from, \eg, semileptonic decays of \PB hadrons, are misidentified as signal-like leptons.
These types of background are labeled as nonprompt,
and are estimated by means of simulated samples.
Multiboson and \ttbar production in association with a gauge boson are rare SM processes that can also contribute to the SRs,
and their contributions are obtained from simulation as well.

To correct the mismodeled \pt distribution of the \PZ boson in the DY simulation~\cite{Sirunyan:2019bzr,Lindert:2017olm},
we apply a $K$ factor as a function of the generator-level \PZ boson \pt (``$\pt(\PZ)$ correction").
The NLO-to-LO $K$ factor in QCD is obtained by taking the ratio between DY samples simulated to NLO and LO precision.
The uncertainties from renormalization and factorization scale, PDF variations, and the statistical uncertainties in the MC samples are taken into account.
The nNLO electroweak correction (combination of the $\mathcal{O}(\alpha^{2}\alpS)$ contributions and the $\mathcal{O}(\alpha^{3}\alpS)$ electroweak Sudakov logarithms in the next-to-leading log approximation) is taken from Ref.~\cite{Lindert:2017olm}.
We define DY CRs to be the same as the SR but modify the dilepton mass requirement to $60 < \mll < 150\GeV$ (blue region in Fig.~\ref{fig:BkgdSchem}).
For the boosted DY CRs, a lepton is not required to be located inside an AK8 jet, and the \LSFiii requirement is removed when collecting the AK8 jets.
Owing to a limitation of the DY MC generator, which is restricted to the LO simulation of events with a maximum of four partons in the hard scattering,
a shape discrepancy is observed in the leading AK4 and AK8 jet \pt distributions in the resolved and boosted DY CRs, respectively.
We correct the shape of the jet \pt distributions to match with those observed in data by a bin-by-bin rescaling (``DY reshape").
Since this mismodeling is due to the higher-order effect of QCD jet splitting, the dielectron and dimuon events will be affected in a similar way, thus the two event samples are summed when extracting the reweighting values, typically ranging from 0.8 to 1.2.
The DY reshape correction modifies the shape of the \mlljj and \mlJ distributions and brings the distribution in simulation into good agreement with data.
The uncertainty assigned to the DY reshape correction is dominated by the statistical uncertainty, and is treated as uncorrelated across bins in the invariant mass spectrum.
The normalization of the DY processes is allowed to vary in the fitting procedure, which is described in Section~\ref{sec:result}.
Several checks were performed to validate the extrapolation from low to high \mll, including verifying the agreement in the intermediate region, $150 < \mll < 400(200)\GeV$ for the resolved (boosted) analysis.

To constrain the normalization of the \ttbar and \tW backgrounds in each SR, a corresponding control region (flavor CR) dominated by \ttbar and \tW events is defined by requiring the two leptons to have different flavors (the green region in Fig.~\ref{fig:BkgdSchem}).
For the boosted analysis, separate control regions are used for the dielectron and dimuon searches.
These require a lead muon plus a jet with an electron contained within it (\Pe-jet), and a lead electron plus a jet with a muon contained within it (\PGm-jet), respectively.
The normalizations of these backgrounds are extracted by performing a simultaneous fit across the SRs and flavor CRs.

\begin{figure}[htbp]
  \centering

  \includegraphics[width=0.49\textwidth]{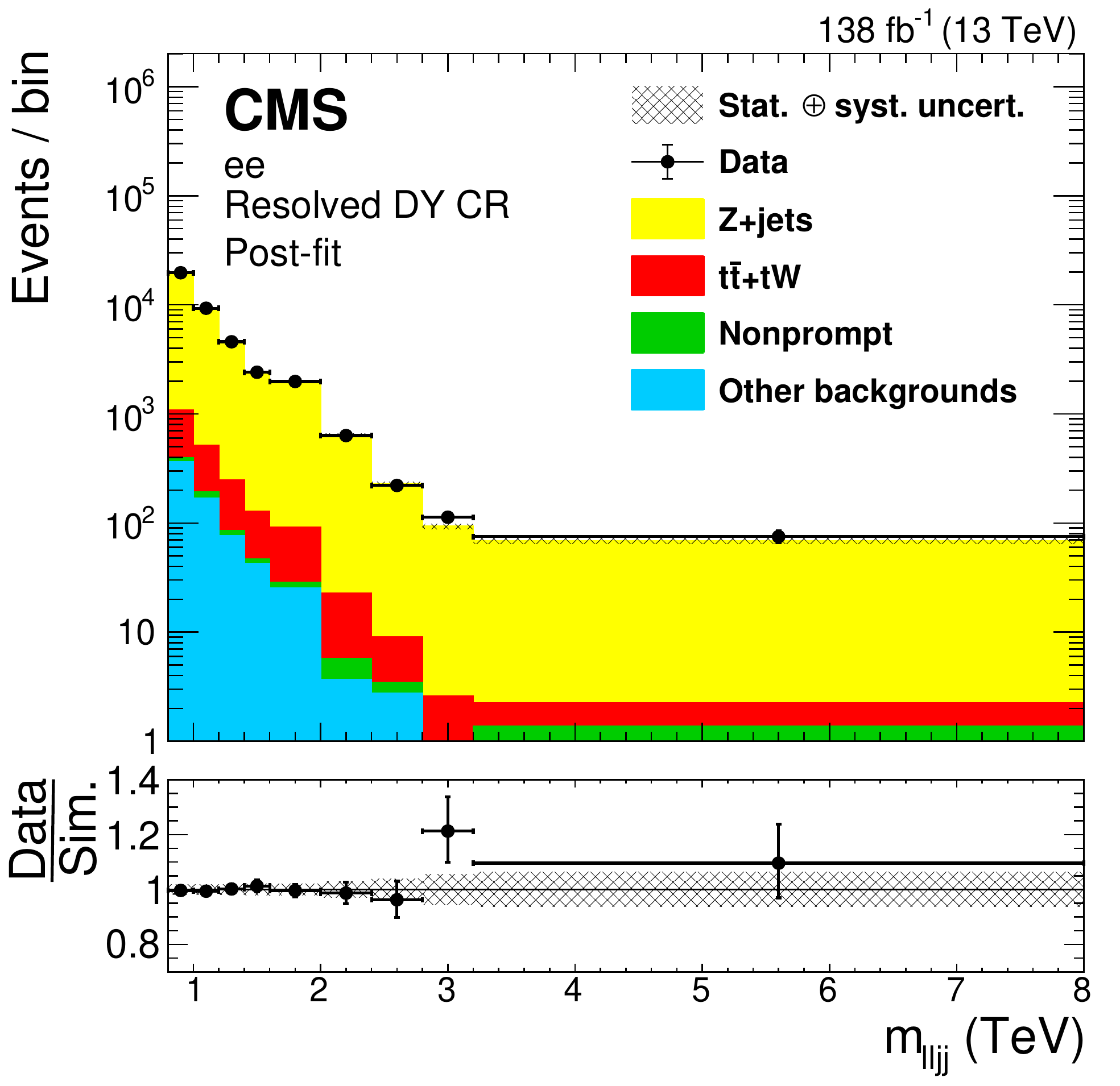}
  \includegraphics[width=0.49\textwidth]{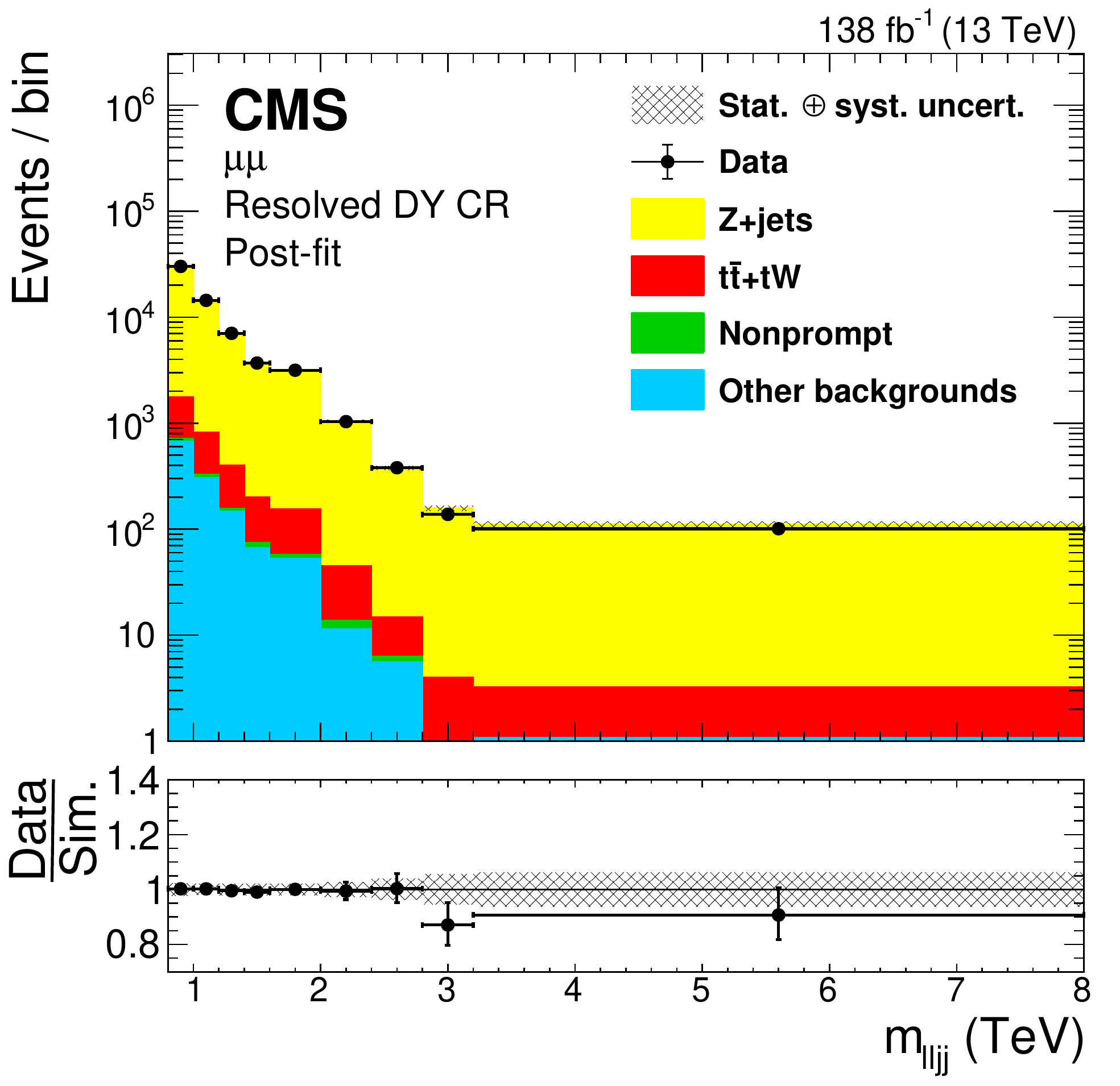}

  \includegraphics[width=0.49\textwidth]{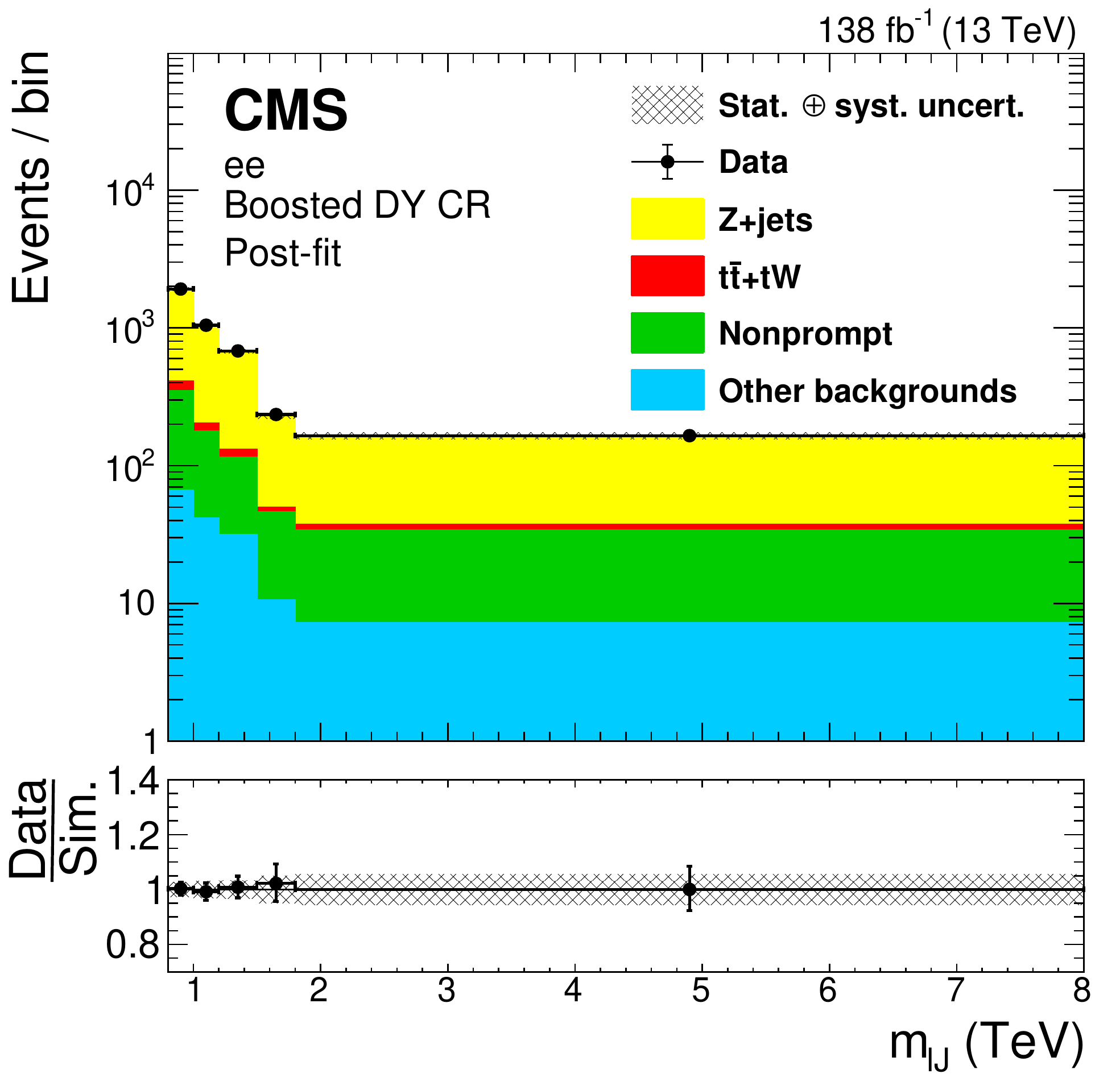}
  \includegraphics[width=0.49\textwidth]{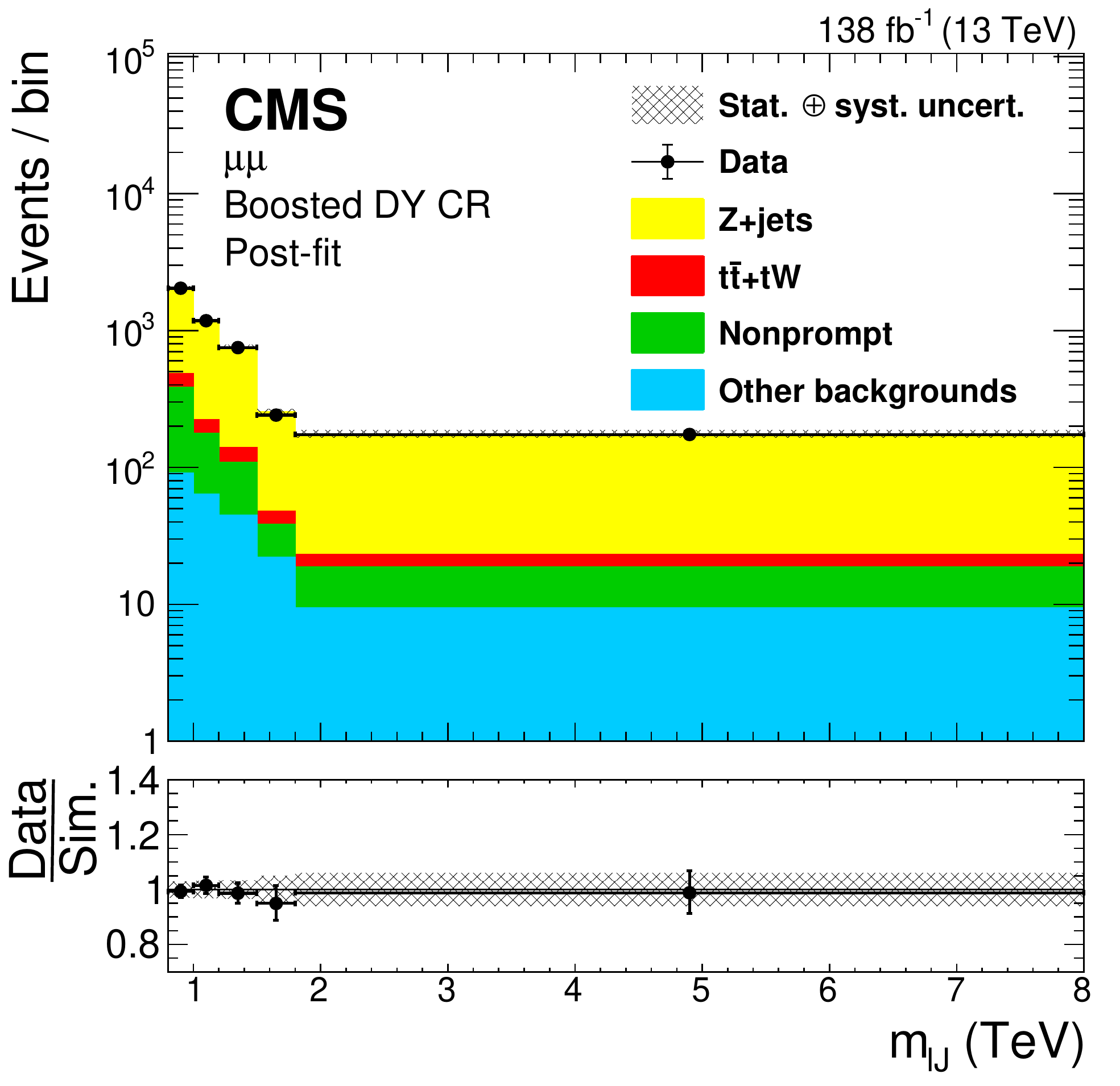}

  \caption{
    The \mlljj (\mlJ) distributions in the resolved (boosted) DY CRs are shown in the upper (lower) row.
    Results in the $\Pe\Pe$ ($\PGm\PGm$) channels are shown in the left (right) plots.
    The hatched uncertainty bands on the simulated background histograms include statistical and systematic components.
  }
  \label{fig:DYCRPlots}
\end{figure}

\begin{figure}[t]
  \centering

  \includegraphics[width=0.49\textwidth]{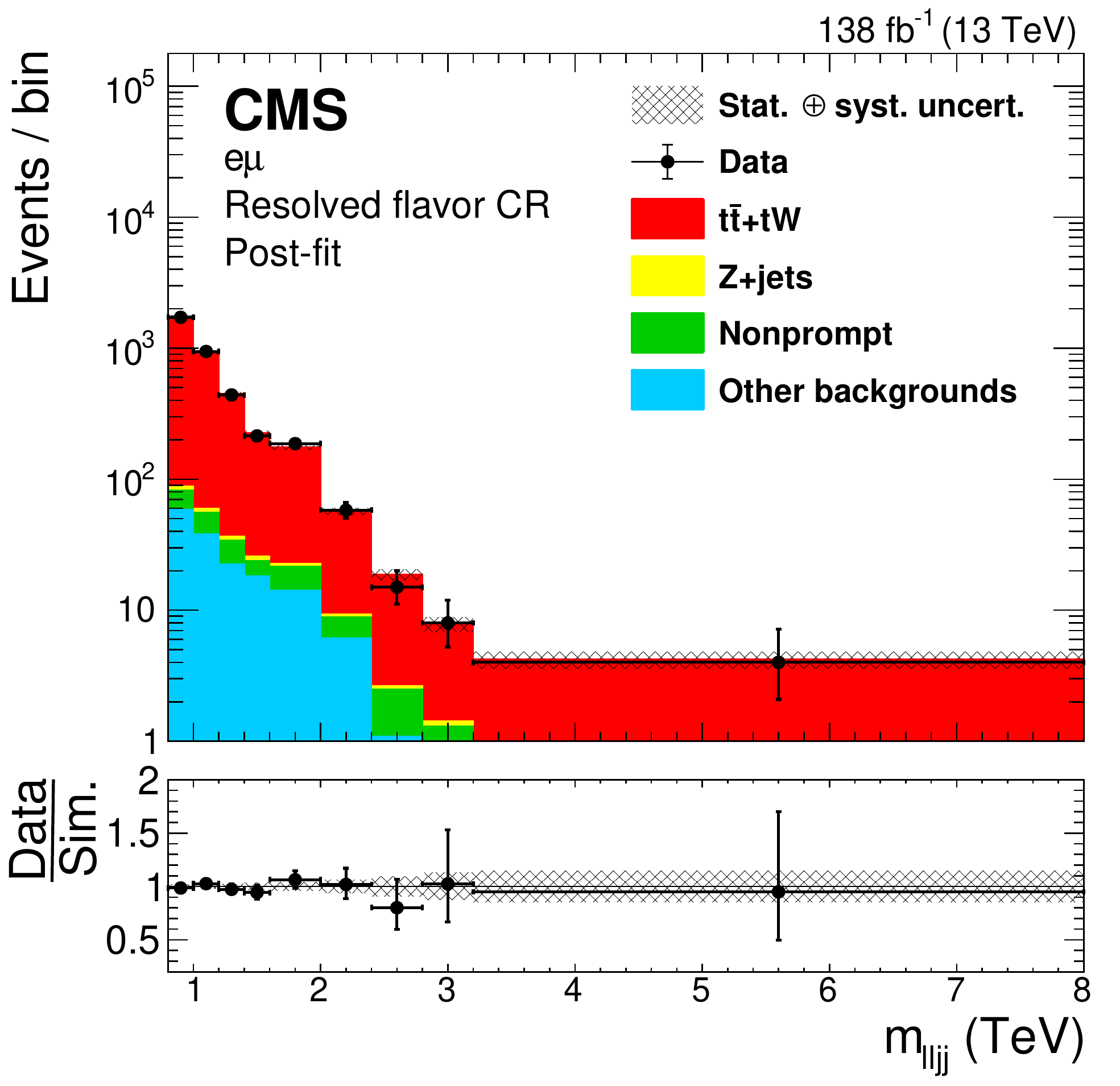}

  \includegraphics[width=0.49\textwidth]{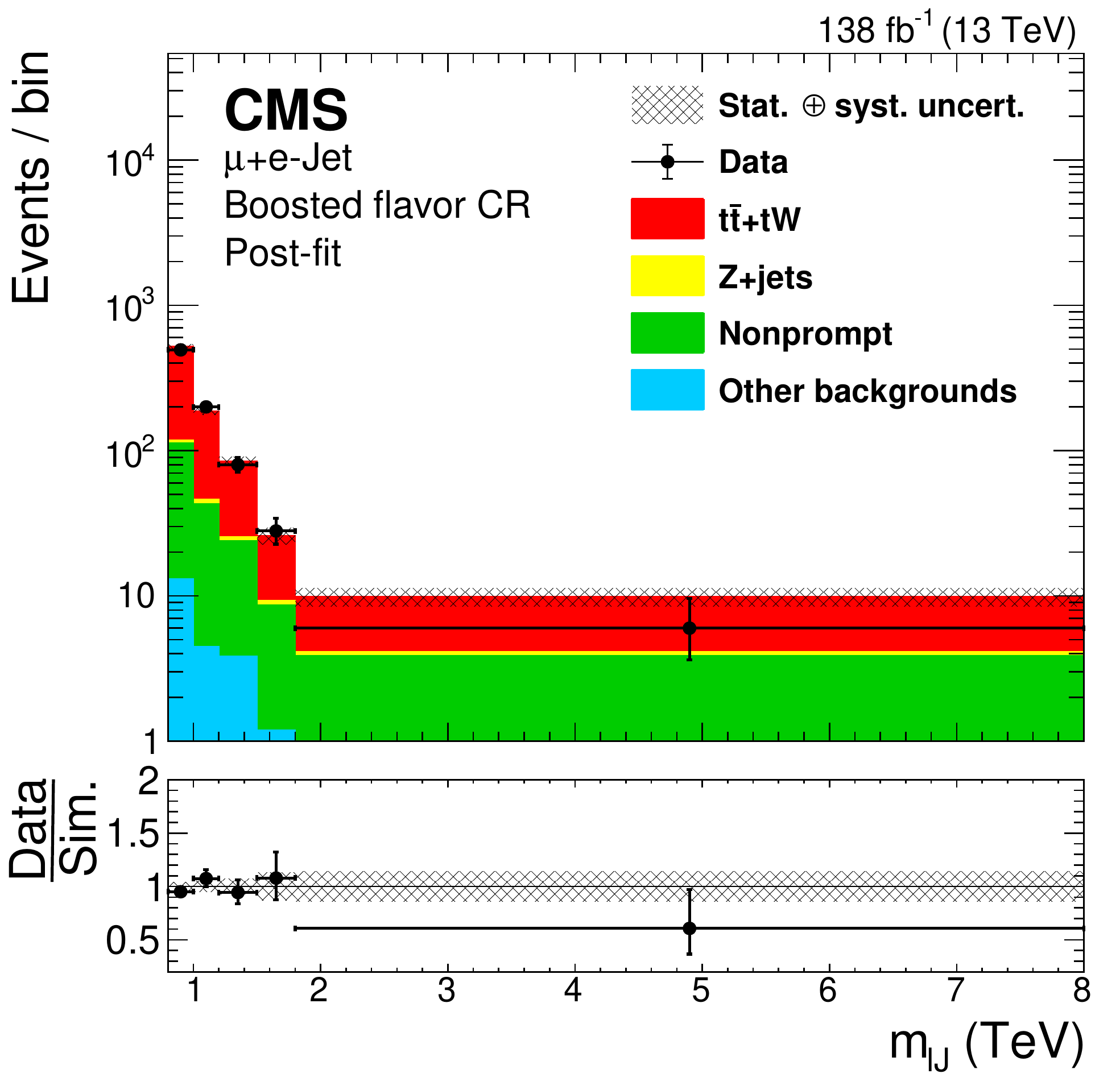}
  \includegraphics[width=0.49\textwidth]{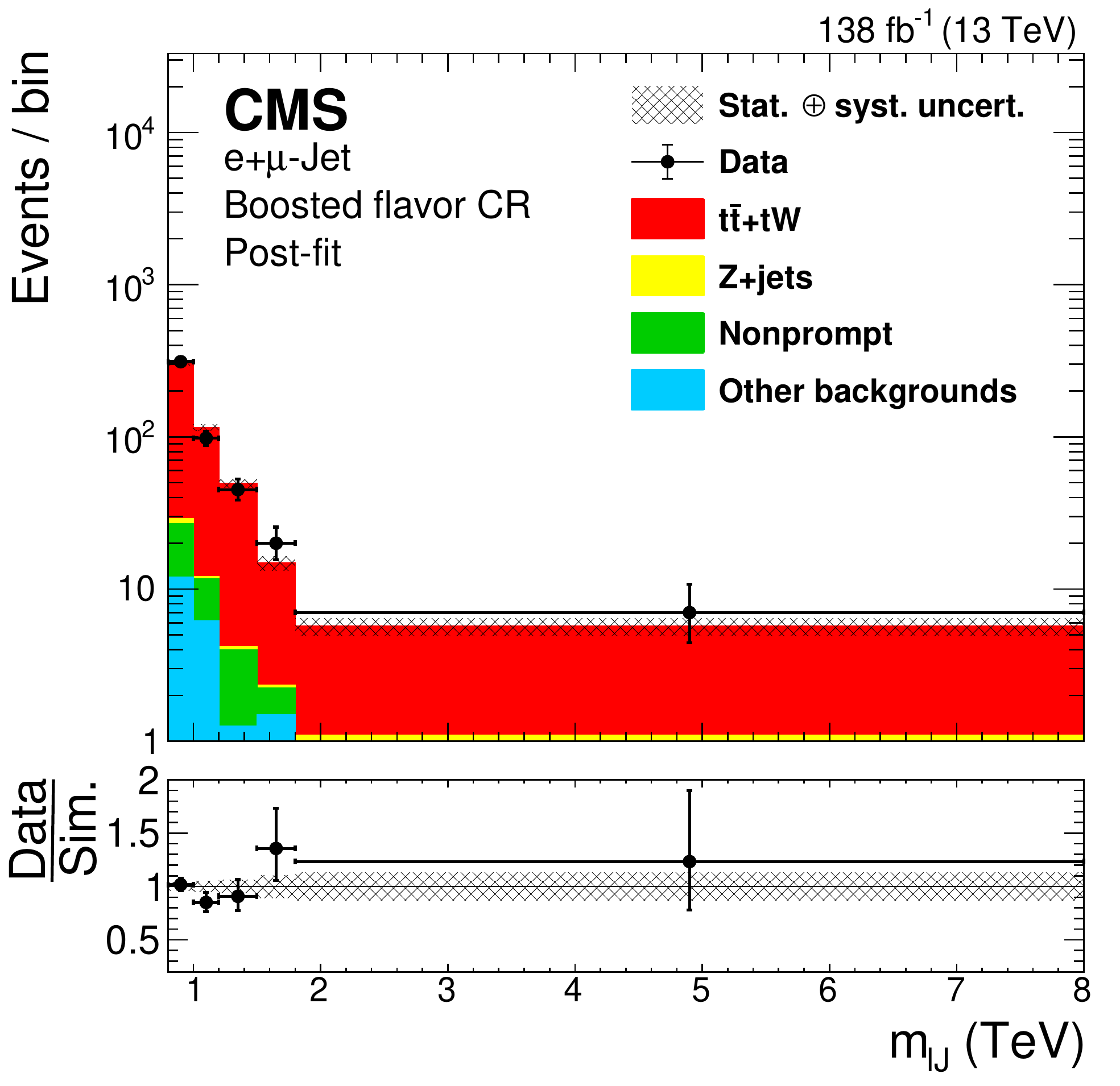}

  \caption{
    The reconstructed mass of the \PWR boson in the resolved (upper), boosted with \Pe-jet (lower left), and boosted with \PGm-jet (lower right) flavor CR.
  }

  \label{fig:EMuCR}
\end{figure}

The invariant mass distributions in the DY CRs, flavor CRs, and the SRs are simultaneously fitted (Section~\ref{sec:result}).
The \mlljj and \mlJ distributions after the fitting (post-fit) in the DY CRs and flavor CRs are shown in Figs.~\ref{fig:DYCRPlots} and~\ref{fig:EMuCR}, respectively.

\section{Systematic uncertainties}
\label{sec:systUnc}

The \mlljj (\mlJ) distribution is used to perform a binned maximum likelihood fit in the resolved (boosted) SR.
The shapes of the invariant mass distributions are affected by a number of systematic uncertainties.
Each systematic uncertainty is treated either as uncorrelated or fully correlated across the three data-taking years.
A complete list of systematic uncertainties and their correlations is given in Table~\ref{tab:systUnc}.

The integrated luminosities for the 2016, 2017, and 2018 data-taking years have 1.2--2.5\% individual uncertainties~\cite{CMS-LUM-17-003,CMS-PAS-LUM-17-004,CMS-PAS-LUM-18-002}, while the overall uncertainty for the 2016--2018 period is 1.6\%. 
The uncertainties in the expected signal and background yields associated with the pileup simulation are estimated by varying the total cross section of the inelastic scattering used in the simulation by $\pm5\%$. The resultant uncertainties are 0.0--0.9\% and 0.2--1.1\% in signal and background yields, respectively.

The lepton trigger, reconstruction, identification, and isolation efficiencies are 
measured in both data and simulation using $\PZ \to \ell^{\pm}\ell^{\mp}$ ($\ell=\Pe$ or \PGm) events 
and data-to-simulation scale factors are applied to all simulation samples to correct any discrepancies.
The uncertainties in the lepton scale factors are evaluated by changing the scale factors by $\pm 1$ standard deviation from their nominal values.
The momenta of leptons are varied in the simulation within their $\pm1$ standard deviation from their nominal value, 
and the differences in the mass distributions are taken as the uncertainties.
The uncertainties in the jet energy scale and resolutions~\cite{Khachatryan:2016kdb} affect the signal and background yields by less than 5\%.
The jet selection efficiency associated with the requirement $\LSFiii>0.75$ is measured by taking a sample of AK8 jets from boosted \PW bosons in \ttbar events where one \PW boson decays leptonically and the other decays hadronically and injecting a simulated lepton (\Pe or \PGm) in the direction of the hadronically decaying \PW boson, to emulate the 3-prong jet topology. 
The data-to-simulation scale factor obtained for the \LSFiii requirement, ranging from 0.95 to 1.05, is then applied to the simulated samples. 
The resulting uncertainty in the yields amounts to less than 10\% for both the signal and background.

As described in Section~\ref{sec:bkgdEst}, the two corrections applied to the DY simulation are varied within their uncertainties
to estimate their impact on the invariant mass distributions.

Owing to the imperfect description of nonprompt leptons in the simulation,
a conservative 100\% normalization uncertainty, cross-checked with a dedicated control region with zero b jets, is assigned to the nonprompt background estimation.
A 50\% uncertainty is assigned to the rare SM background.
The theoretical uncertainties originating from the strong coupling constant \alpS, PDFs, and renormalization/factorization scales are the dominant sources of uncertainties for the signal shape estimate.
The PDF and \alpS uncertainties are estimated from the standard deviation of the weights from the PDF replicas provided in the NNPDF3.1 PDF set~\cite{Ball:2017nwa}, following the PDF4LHC procedure~\cite{Butterworth:2015oua}.

\begin{table}[htb]
  \centering
  \topcaption{
Summary of the relative uncertainties in the total yields of the signal and background (bkgd) predictions.
The uncertainties are given for the resolved (boosted) SR.
The values for the signal correspond to $\mWR=5\TeV$.
The range given for each systematic uncertainty source covers the variation across the years and the year-to-year treatment is shown as either fully correlated (C) or uncorrelated (U).
The EW1, EW2, and EW3 entries are the uncertainties from the $\mathcal{O}(\alpha^{2})$ Sudakov terms, the NLO portion of the nNLO EW uncertainty, and the uncertainty in the Sudakov approximation at nNLO, respectively.
  }
  \cmsTable{
  \begin{tabular}{lcccccc}
\hline
\multirow{2}{*}{Source} & \multirow{2}{*}{Process} &\multirow{2}{*}{Corr.} & $\Pe\Pe$ bkgd. & $\Pe\Pe$ signal & $\PGm\PGm$ bkgd. & $\PGm\PGm$ signal \\
                        &                                       &     &  (\%)           & (\%)            & (\%)           & (\%) \\
\hline
Integrated luminosity & All & C & 1.6 (1.6) & 1.6 (1.6) & 1.6 (1.6) & 1.6 (1.6) \\
Electron reconstruction & All & C & 1.0--1.6 (0.5--0.8) & 0.8--1.4 (0.4--0.8) & \NA & \NA \\
Electron energy resolution & All & C & $<$0.1 ($<$0.1) & $<$0.1 ($<$0.1) & \NA & \NA \\
Electron energy scale & All & C & 0.5--1.9 (0.5--2.4) & 0--0.2 (0--0.4) & \NA & \NA \\
Electron identification & All & C & 3.1--3.3 (1.8--1.9) & 4.1--4.4 (2.1--2.4) & \NA & \NA \\
Electron trigger & All & U & 0--0.1 (0.2--0.4) & $<$0.1 (0.1--0.2) & \NA & \NA \\
Muon reconstruction & All & C & \NA & \NA & 0.4--1.0 (0.3--0.7) & 4.4--36.8 (5.6--30.7) \\
Muon momentum scale & All & C & \NA & \NA & 0.4--2.5 (0.5--3.7) & 0.1--0.2 (0.1--0.3) \\
Muon identification & All & C & \NA & \NA & 0.2--1.2 (0.1--0.6) & 0.2--1.1 (0.1--0.5) \\
Muon isolation & All & C & \NA & \NA & 0.1--0.2 (0--0.1) & 0.1--0.2 (0--0.1) \\
Muon trigger & All & U & \NA & \NA & 0.1--0.2 (0.1--0.2) & 0.7--1.6 (0.5--1.3) \\
Jet energy scale & All & C & 1.9--4.2 (0.9--1.9) & 0--0.2 (0--0.2) & 2.1--3.5 (0.6--0.9) & 0--0.2 (0--0.4) \\
Jet energy resolution & All & U & 0.5--1.5 (0.7--2.0) & 0--0.3 (0--0.4) & 0.2--1.3 (0.2--1.1) & 0--0.3 (0--0.3) \\
Jet mass scale & All & C & $<$0.1 (1.0--1.8) & $<$0.1 (0.1--1.0) & $<$0.1 (1.4--1.7) & $<$0.1 (0.2--1.0) \\
LSF scale factor & All & U & \NA~(6.7--8.7) & \NA~(6.7--8.7) & \NA~(5.8--7.1) & \NA~(5.8--7.1) \\
Pileup modeling & All & C & 0.2--1.1 (0.5--1.2) & 0.1--0.8 (0.1--0.9) & 0.3--0.5 (0.2--1.2) & 0.1--0.5 (0--0.6) \\
\pt(\PZ) correction, MC stat. & \DYJ & C & 0.6--2.0 (0.6--2.0)& \NA & 0.6--2.0 (0.6--2.0)& \NA \\
\pt(\PZ) correction, renorm./fact. scales & \DYJ & C & 6.3--7.1 (6.7--7.3)& \NA & 6.1--7.1 (6.8--7.4)& \NA \\
\pt(\PZ) correction, PDF replicas & \DYJ & C & 0.6--2.0 (0.6--2.0)& \NA & 0.6--2.0 (0.6--2.0)& \NA \\
\pt(\PZ) correction, PDF \alpS & \DYJ & C & 0.8--4.7 (0.8--5.2)& \NA & 0.8--4.7 (0.8--5.3)& \NA \\
\pt(\PZ) correction, EW1~\cite{Lindert:2017olm} & \DYJ & C & $<$0.1 ($<$0.1)& \NA & $<$0.1 ($<$0.1)& \NA \\
\pt(\PZ) correction, EW2~\cite{Lindert:2017olm} & \DYJ & C & 0.3 (0.3)& \NA & 0.3 (0.3)& \NA \\
\pt(\PZ) correction, EW3~\cite{Lindert:2017olm} & \DYJ & C & 0.1 (0.1)& \NA & 0.1 (0--0.1)& \NA \\
DY reshape & \DYJ & C & 8.5--9.1 (10.1--11.6)& \NA & 8.5--9.2 (9.7--11.6)& \NA \\
Nonprompt background normalization & Nonprompt & U & 100 (100) & \NA & 100 (100) & \NA  \\
Rare SM background normalization & Others & C & 50 (50) & \NA & 50 (50) & \NA  \\
PDF replicas & Signal & C & \NA & 5.9--11.9 (8.8--40.3) & \NA & 2.8--6.8 (17.5--40.6) \\
\alpS & Signal & C & \NA & 0--0.2 (0.2--1.3) & \NA & 0--0.2 (0.2--1.2) \\
Renorm./fact. scales & Signal & C & \NA & 0--0.1 (0.3--2.3) & \NA & 0--0.1 (2.1--2.9) \\
\hline
  \end{tabular}
  }
  \label{tab:systUnc}
\end{table}

\section{Results}
\label{sec:result}

\begin{figure}[htb!]
  \centering

  \includegraphics[width=0.49\textwidth]{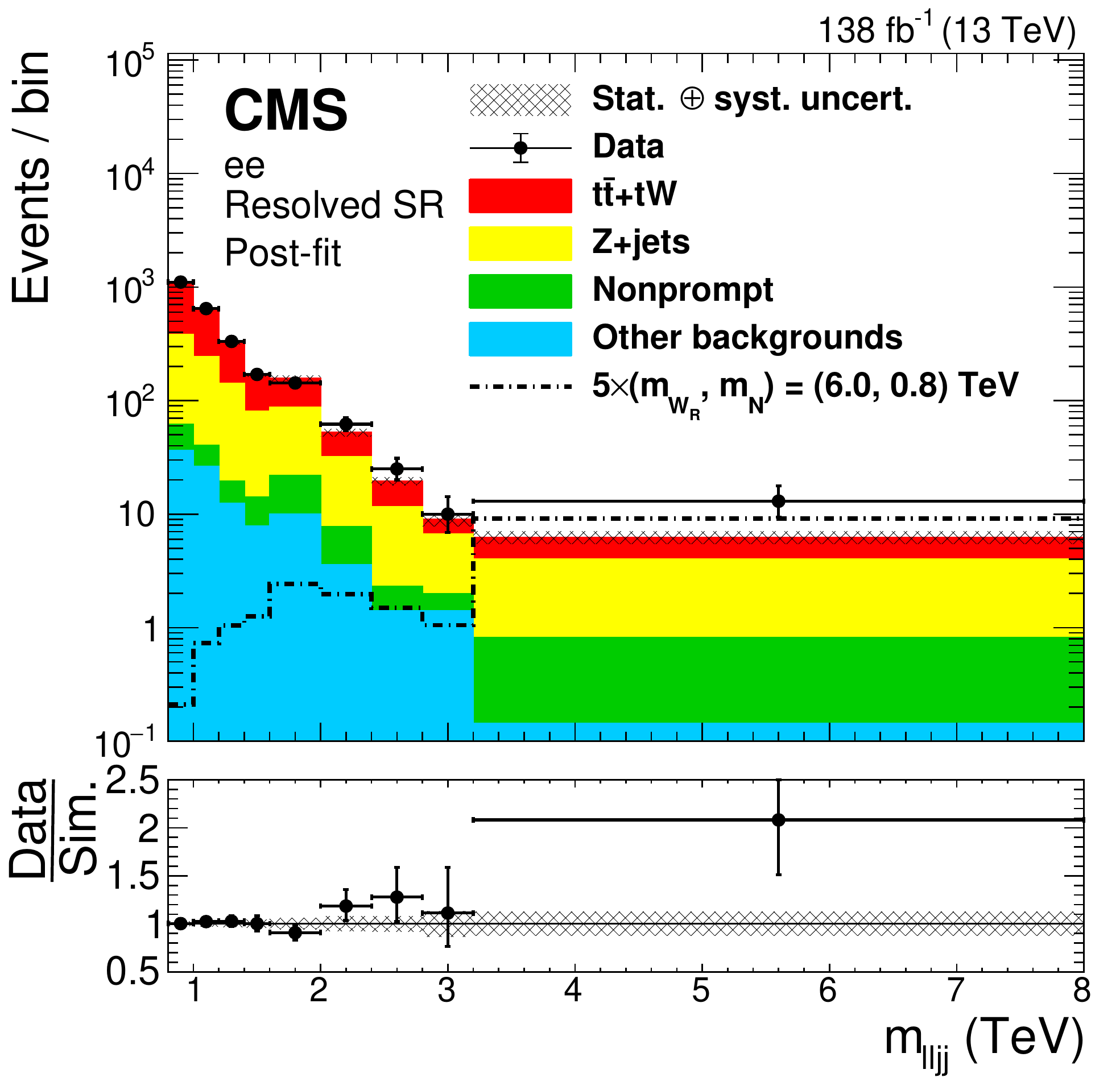}
  \includegraphics[width=0.49\textwidth]{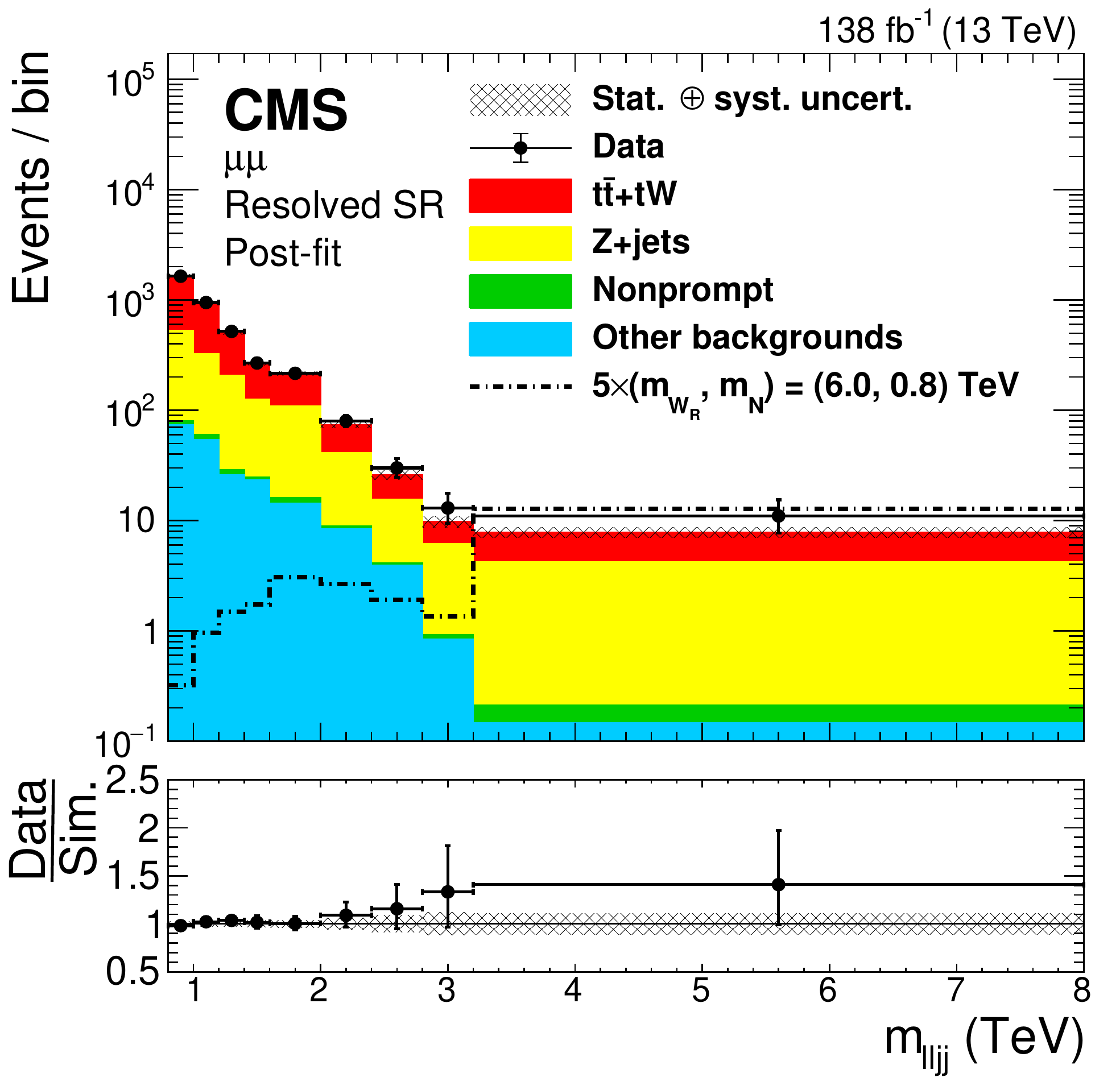}

  \includegraphics[width=0.49\textwidth]{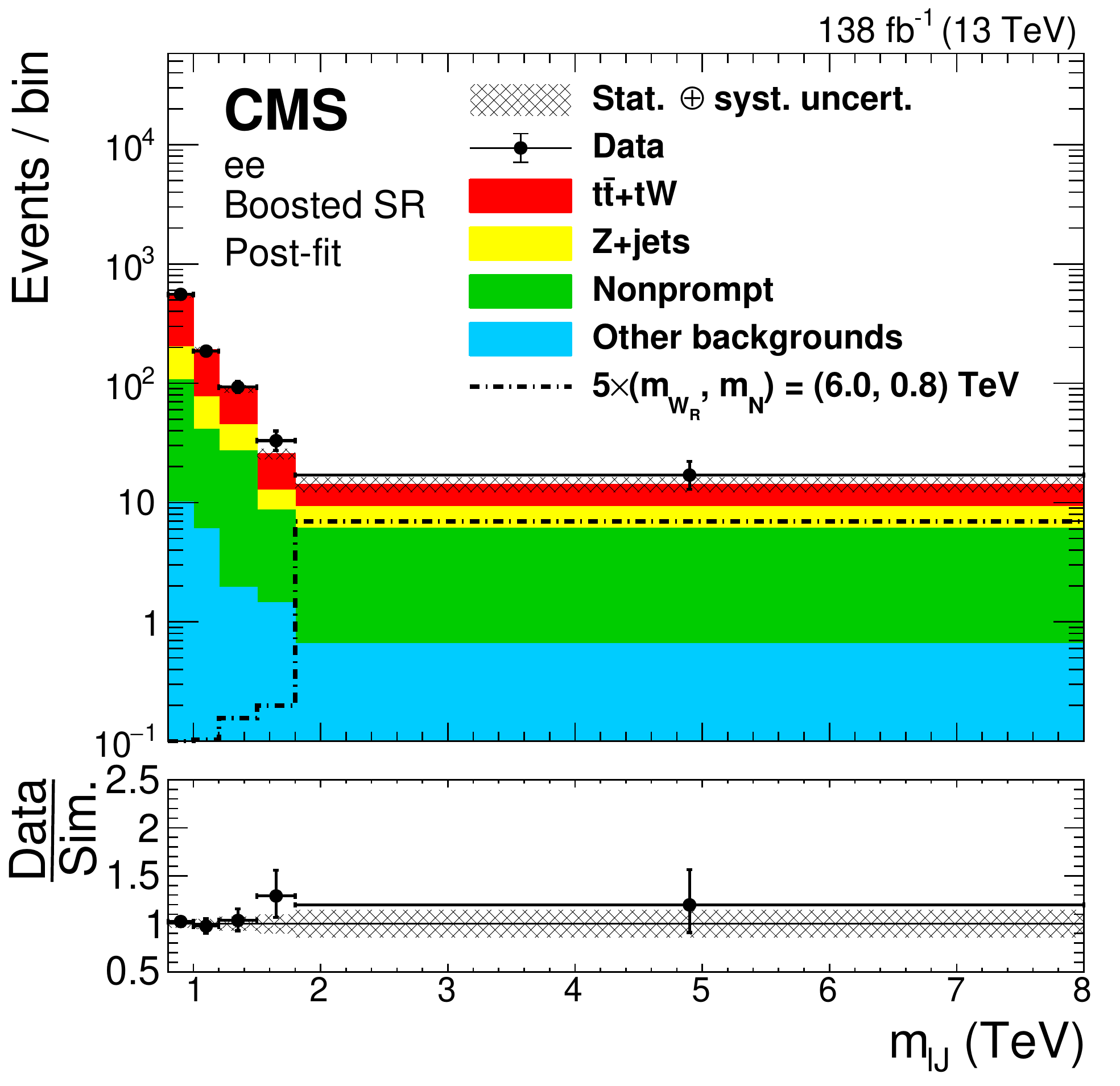}
  \includegraphics[width=0.49\textwidth]{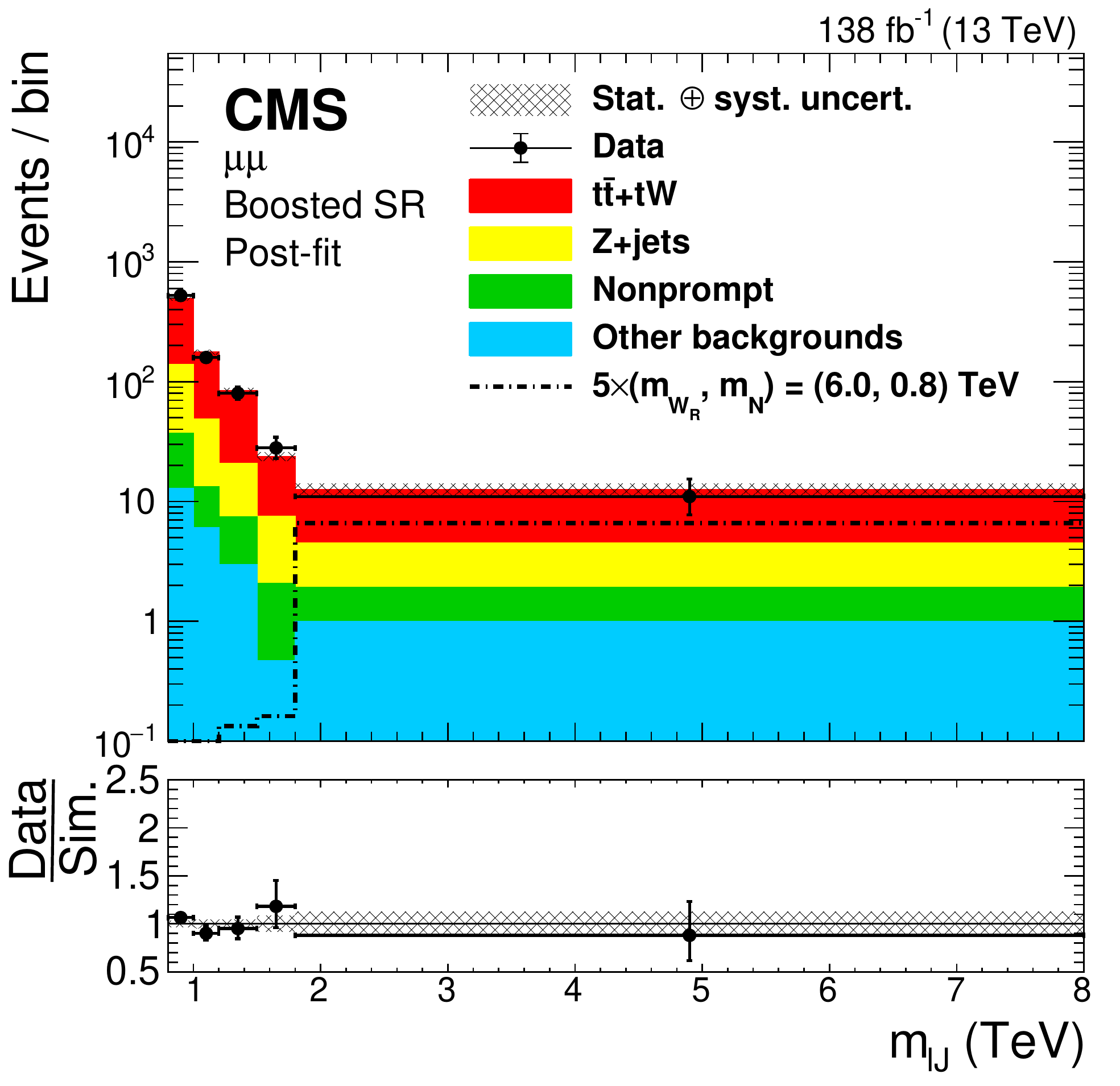}

  \caption{
    The background-only post-fit \mlljj (\mlJ) distributions in the resolved (boosted) SR are shown in the upper (lower) plot.
    Results for the dielectron (dimuon) channel are shown on the left (right).
    Statistical and systematic uncertainties in the expected background yields are represented by the hatched band.
    The simulated signal distribution is scaled up by a factor of five to enhance visibility.
  }
  \label{fig:YearCombinedSR}
\end{figure}

\begin{figure}[htb!]
  \centering

  \includegraphics[width=0.49\textwidth]{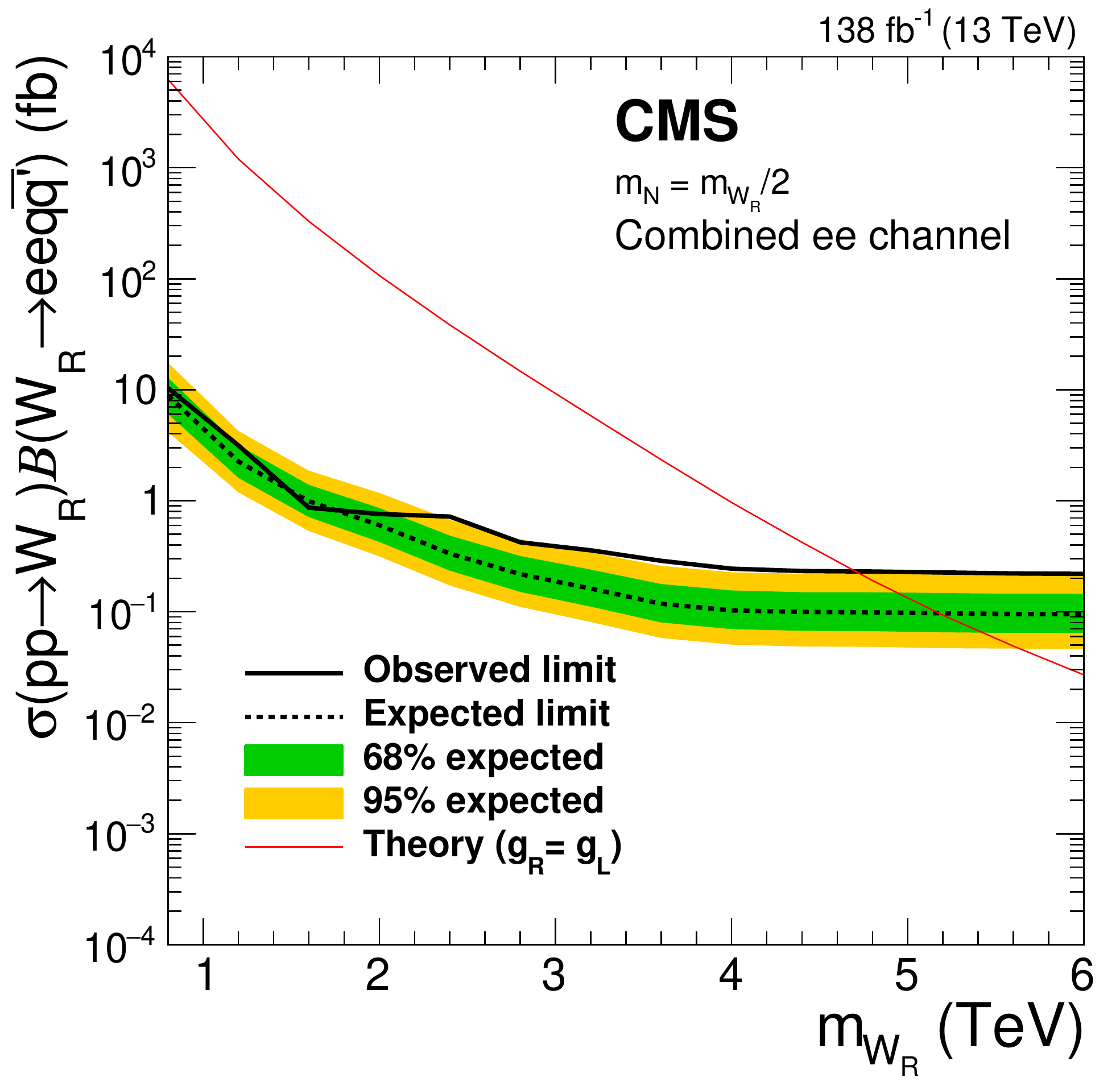}
  \includegraphics[width=0.49\textwidth]{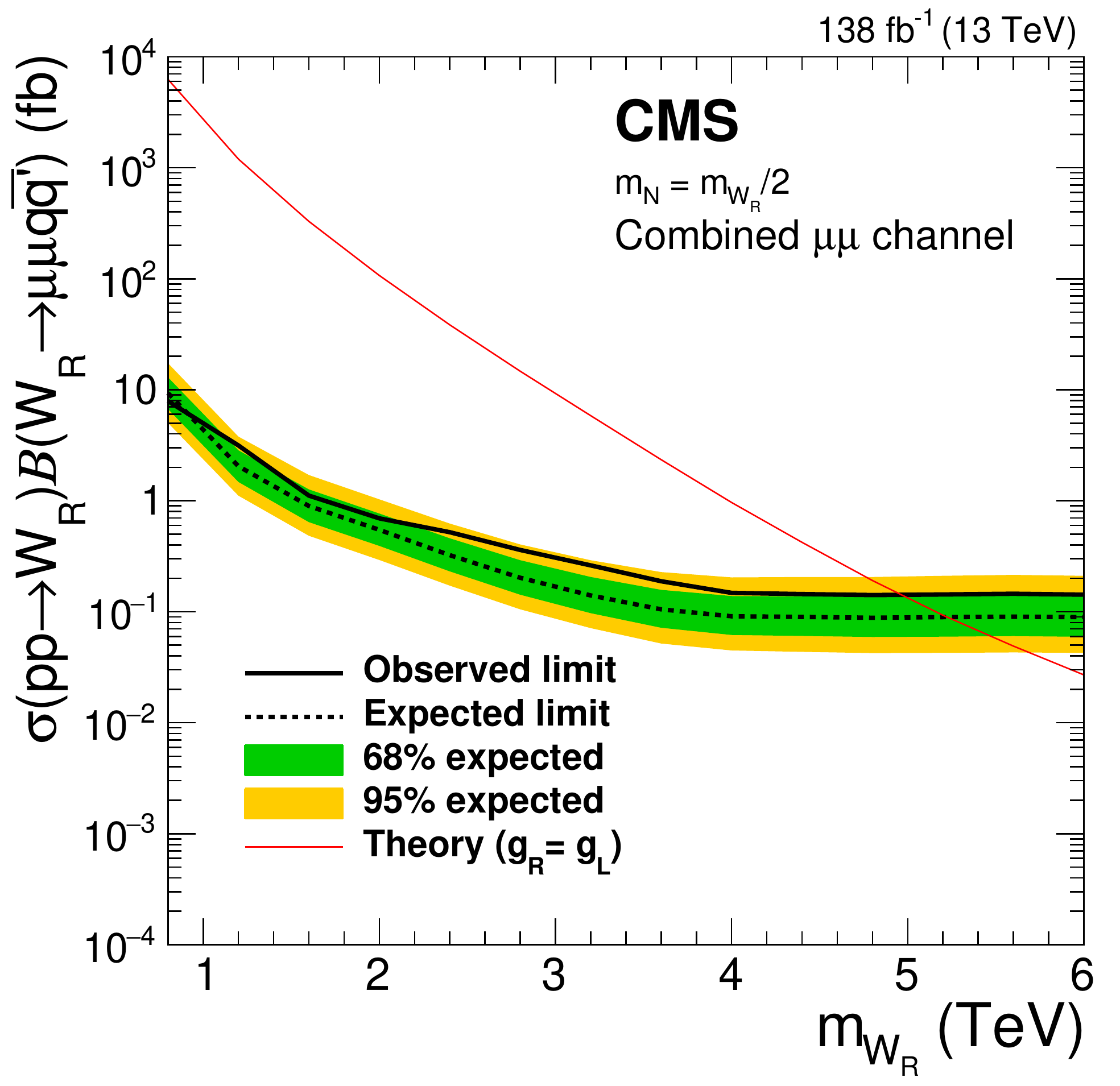}

  \includegraphics[width=0.49\textwidth]{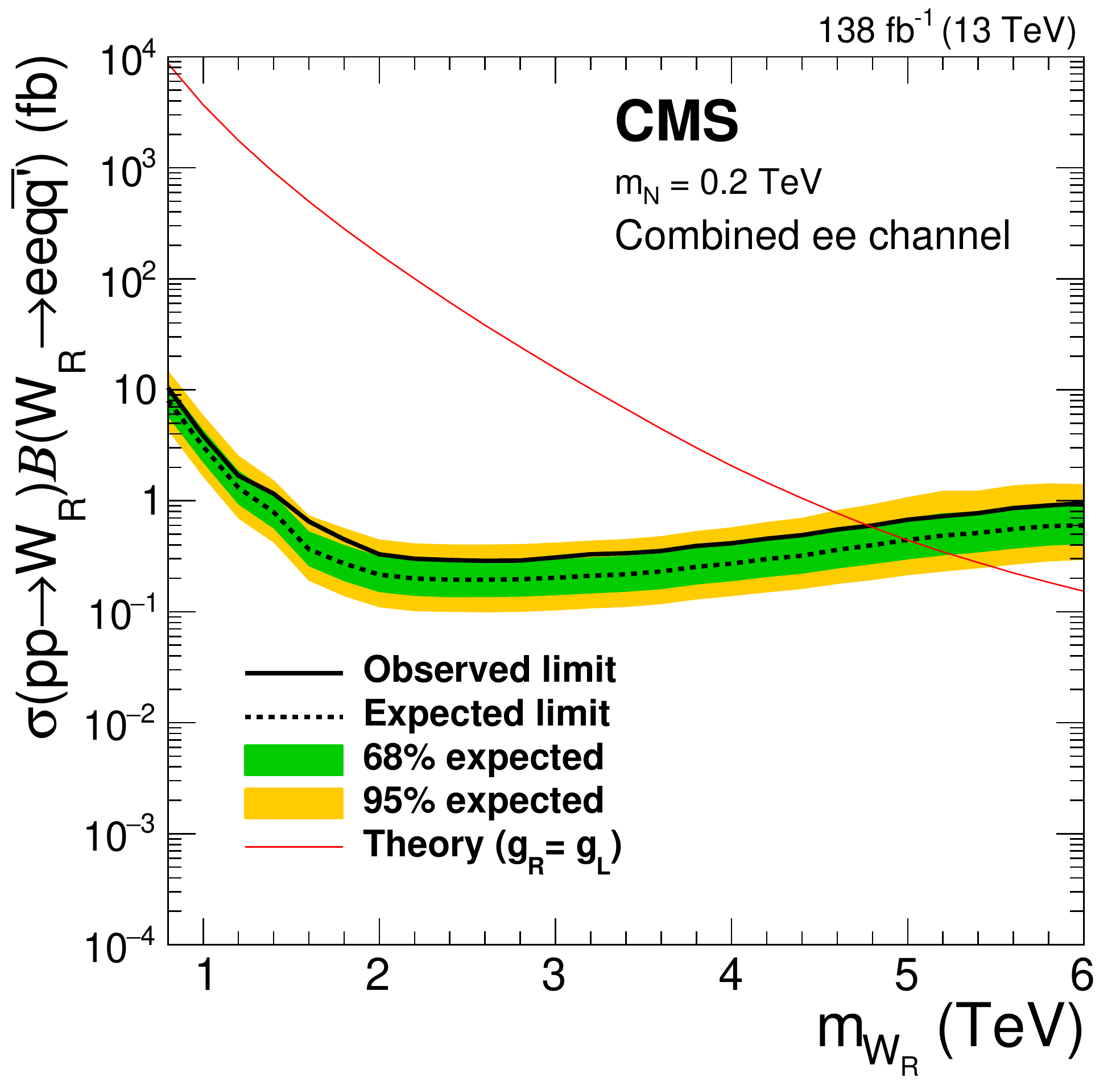}
  \includegraphics[width=0.49\textwidth]{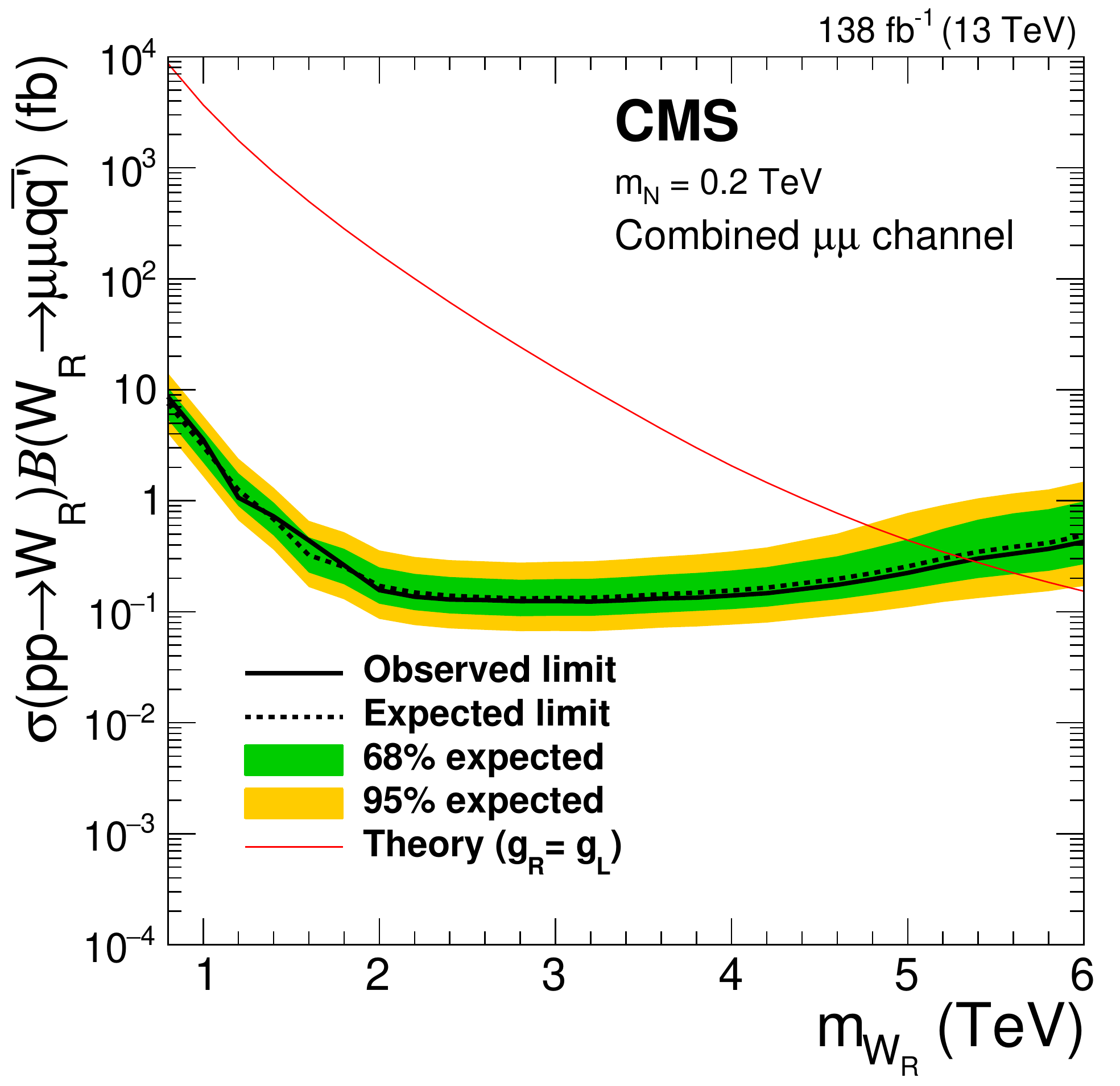}

  \caption{
    The expected (black dashed line) and the observed (black solid line) 95\% \CL upper limits on the product of the cross section for \PWR production and the branching fractions for the electron channel (left) and muon channel (right) from the combination of the resolved and boosted categories.
    The plots in the upper (lower) row are the results for the $\mN = \mWR/2$ ($\mN=0.2\TeV$) mass point, which correspond to the resolved (boosted) \PWR topology.
    The green (inner) and yellow (outer) bands indicate the 68 and 95\% coverage of the expected upper limits.
    The red solid lines represent the values expected from the theory~\cite{Mattelaer:2016ynf}.
  }
  \label{fig:1DFullRun2Limits}
\end{figure}

\begin{figure}[htbp!]
  \centering

  \includegraphics[width=0.49\textwidth]{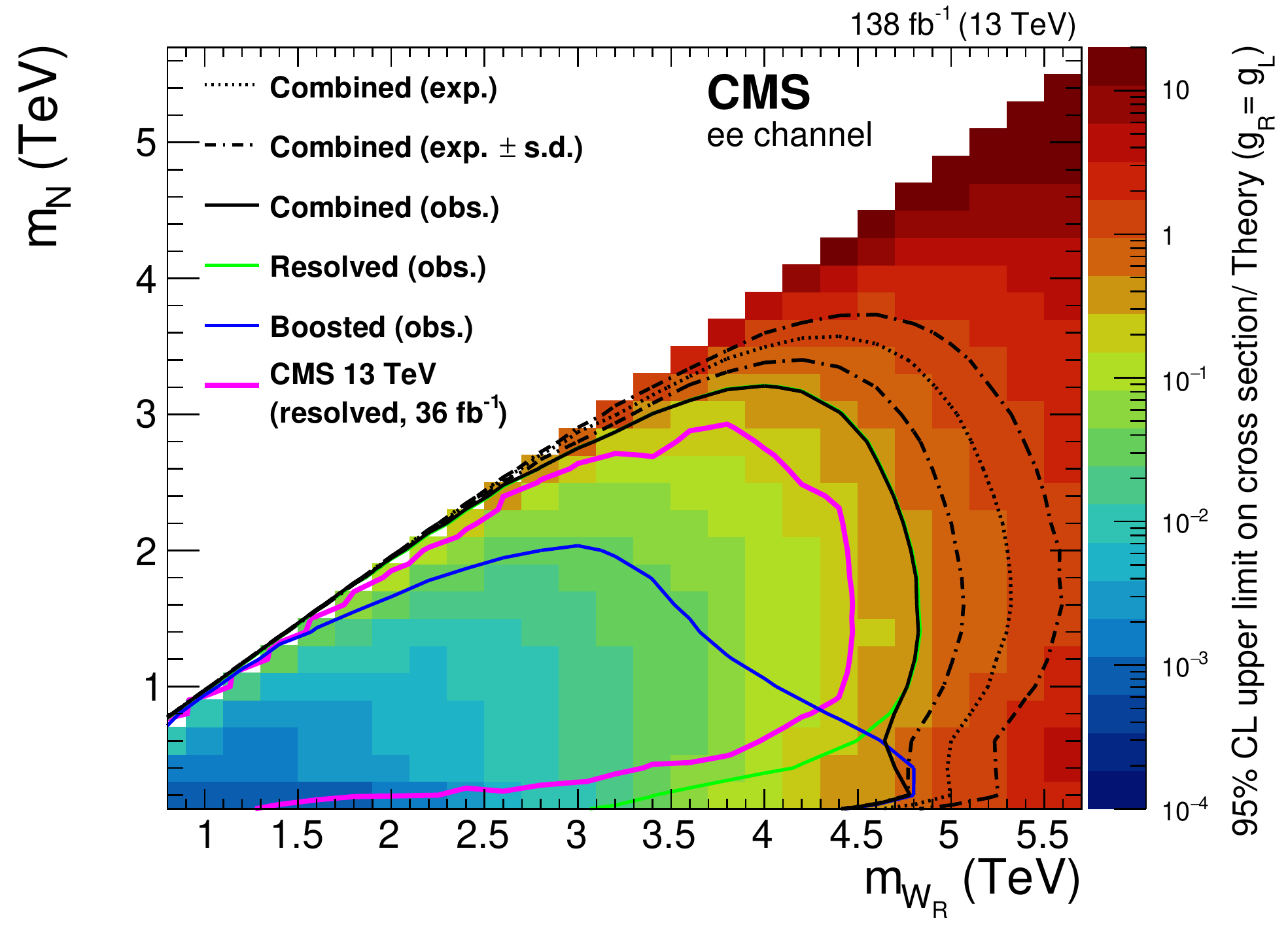}
  \includegraphics[width=0.49\textwidth]{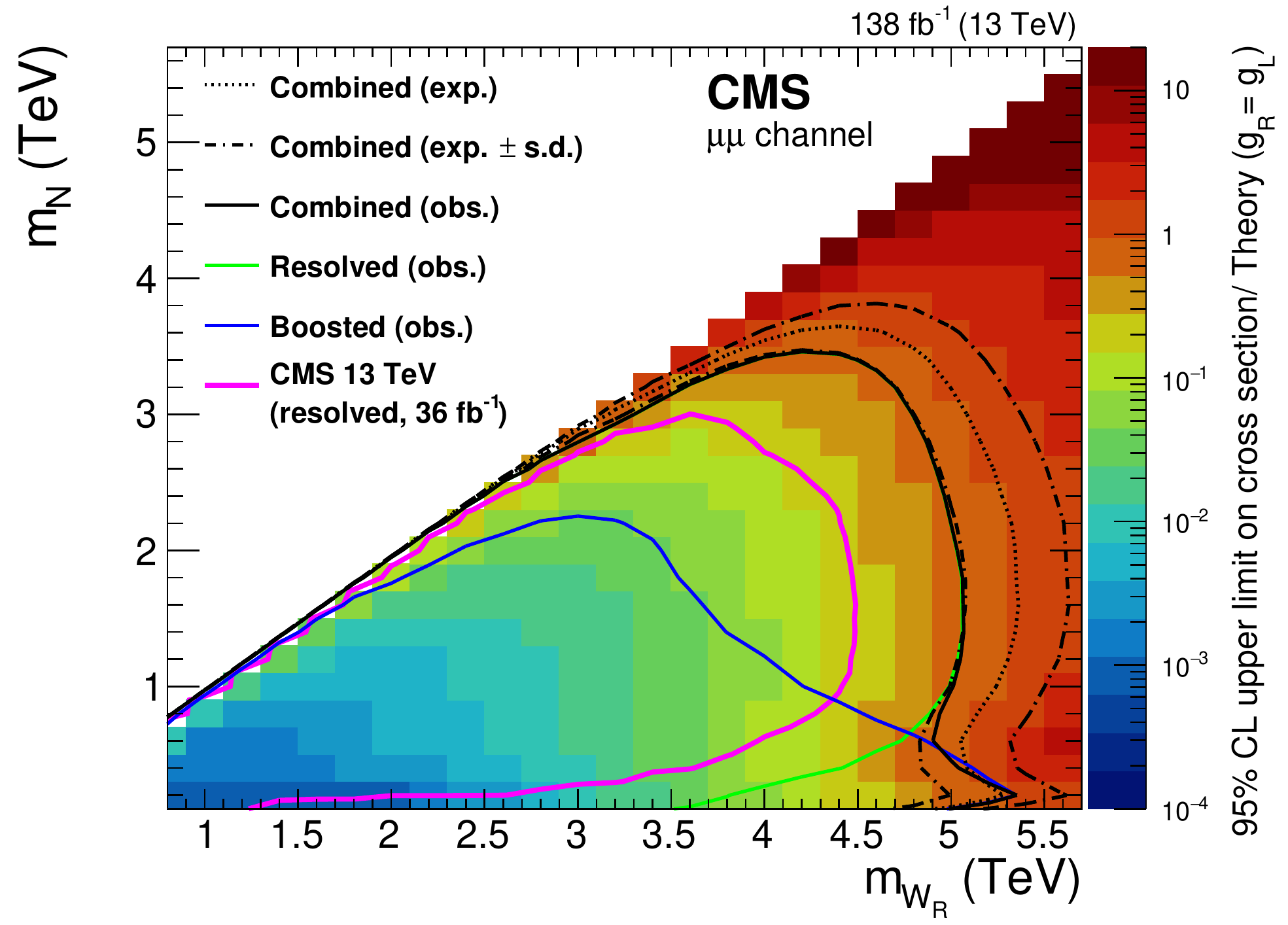}

  \caption{
    The observed 95\% \CL upper limits on the product of the production cross sections and the branching fractions of a right-handed \PWR boson divided by the theory expectation for a coupling constant \gR equal to the SM coupling of the \PWR boson (\gL), for the electron channel (left) and muon channel (right).
    The observed exclusion regions are shown for the resolved (solid green), boosted (solid blue), and combined (solid black) channels, together with the expected exclusion region for the combined result (dotted black). 
    The dash-dotted lines represent the 68\% coverage of the boundaries of the expected exclusion regions. 
    The observed exclusion regions obtained in the previous search performed by the CMS Collaboration~\cite{Sirunyan:2018pom} are bounded by the magenta lines.
    The biggest improvement can be seen in the $\mN < 0.5\TeV$ region, where the new boosted category greatly improves the sensitivity over the previous result.
  }
  \label{fig:2DFullRun2Limits}
\end{figure}

A maximum-likelihood fit is performed on the \mlljj (\mlJ) distributions in the resolved (boosted) SRs and CRs, with the systematic uncertainties described in Section~\ref{sec:systUnc} treated as nuisance parameters.
The background-only post-fit invariant mass distributions are shown in Fig.~\ref{fig:YearCombinedSR} and no significant excess over the background expectations is observed.
A simulated signal distribution corresponding to mass values at the limit of sensitivity is shown for comparison. 
The shape of this distribution is very similar for the resolved and the boosted analysis, with the peak falling in the last bin. 
The enhancement visible around 2\TeV in the resolved distribution is due to sculpting of the \PWR off-shell mass distribution.
The upper limits on the product of the cross section for \PWR production and the branching fractions $\sigma (\Pp\Pp \to \PWR) \mathcal{B}(\PWR \to \Pe\Pe(\PGm\PGm)\cPq\overline{\cPq'})$ 
for various \mWR and \mN hypotheses are obtained 
using the distributions of the likelihood ratio calculated using the asymptotic approximation~\cite{Cowan:2010js} and the \CLs criterion~\cite{CLS2, CLS1}.
The excluded phase space as a function of \mWR obtained from the expected and observed upper limits at 95\% confidence level (\CL) is shown in Fig.~\ref{fig:1DFullRun2Limits}.
With $\mN=\mWR/2$, the observed (expected) lower limit at 95\% \CL on the mass of the \PWR is 
4.7 (5.2)\TeV and 5.0 (5.2)\TeV for the electron and muon channels, respectively;
for $\mN=0.2\TeV$, limits exclude the phase space up to $\mWR=4.8\,(5.0)$ and $5.4\,(5.3)\TeV$ for the electron and muon channels. 
The local $p$-value of the signal strength, as a function of \mWR and \mN, is obtained from fits to the data with the signal stength at each point treated as a free parameter. 
We observe the most extreme $p$-value to occur in the electron channel for the $(\mWR, \mN) = (6.0, 0.8)\TeV$ mass point and to be $1.58\ten{-3}$, corresponding to a local significance of $2.95\sigma$.
The look-elsewhere effect~\cite{Gross_2010} is taken into account by using pseudo-experiments to calculate the probability under the background-only hypothesis of observing a similar or larger excess in the electron channel across the full analysis mass range.
This probability is $2.7\ten{-3}$, corresponding to a global significance of $2.78\sigma$.
The upper limits across the entire $(\mWR,\mN)$ plane are shown in Fig.~\ref{fig:2DFullRun2Limits}.
The results are compared with previous searches for \PWR bosons and heavy neutrinos performed by the ATLAS Collaboration~\cite{Aaboud:2018spl} using LHC $\Pp\Pp$ collision data at $\sqrt{s}=13\TeV$ and the excluded regions from this analysis in \mWR, with $\mN=\mWR/2$, match these previous results in the electron channel and extend them by 0.3\TeV in the muon channel.
Furthermore, for values of \mN less than 0.5\TeV, the introduction of the boosted analysis provides a significant increase in sensitivity, extending the range of \mWR excluded in this region beyond 4.8 and 5.0\TeV in the electron and muon channels, respectively.

\clearpage

\section{Summary}
\label{sec:summary}

A search for right-handed bosons (\PWR) and heavy right-handed neutrinos (\N) in the left-right symmetric extension of the standard model has been presented.
The analysis is based on proton-proton collision data collected at $\sqrt{s}=13\TeV$ by the CMS detector, corresponding to an integrated luminosity of 138\fbinv.
The final state consists of events with two same-flavor leptons ($\Pe\Pe$ or $\PGm\PGm$) and two quarks, and is identified through two regions: the resolved region, where all four objects are well separated, and the boosted region, where the heavy neutrino decay is identified using jet substructure techniques applied to large-area jets.
The addition of the boosted region greatly improves the search sensitivity in the region where $\mN < 0.5\TeV$.
No significant excess over the standard model background expectations is observed in the invariant mass distributions.
Upper limits are set on the products of the \PWR and \N production cross sections and their branching fraction to two leptons and two quarks assuming that couplings are identical to those of the standard model.
For \N masses \mN equal to half the \PWR mass \mWR ($\mN=0.2\TeV$), \mWR is excluded at 95\% confidence level up to 4.7 (4.8) and 5.0 (5.4)\TeV for the electron and muon channels, respectively.
This analysis provides the most stringent limits on the \PWR mass to date. 

\begin{acknowledgments}
  We congratulate our colleagues in the CERN accelerator departments for the excellent performance of the LHC and thank the technical and administrative staffs at CERN and at other CMS institutes for their contributions to the success of the CMS effort. In addition, we gratefully acknowledge the computing centers and personnel of the Worldwide LHC Computing Grid and other centers for delivering so effectively the computing infrastructure essential to our analyses. Finally, we acknowledge the enduring support for the construction and operation of the LHC, the CMS detector, and the supporting computing infrastructure provided by the following funding agencies: BMBWF and FWF (Austria); FNRS and FWO (Belgium); CNPq, CAPES, FAPERJ, FAPERGS, and FAPESP (Brazil); MES and BNSF (Bulgaria); CERN; CAS, MoST, and NSFC (China); MINCIENCIAS (Colombia); MSES and CSF (Croatia); RIF (Cyprus); SENESCYT (Ecuador); MoER, ERC PUT and ERDF (Estonia); Academy of Finland, MEC, and HIP (Finland); CEA and CNRS/IN2P3 (France); BMBF, DFG, and HGF (Germany); GSRI (Greece); NKFIA (Hungary); DAE and DST (India); IPM (Iran); SFI (Ireland); INFN (Italy); MSIP and NRF (Republic of Korea); MES (Latvia); LAS (Lithuania); MOE and UM (Malaysia); BUAP, CINVESTAV, CONACYT, LNS, SEP, and UASLP-FAI (Mexico); MOS (Montenegro); MBIE (New Zealand); PAEC (Pakistan); MSHE and NSC (Poland); FCT (Portugal); JINR (Dubna); MON, RosAtom, RAS, RFBR, and NRC KI (Russia); MESTD (Serbia); MCIN/AE and PCTI (Spain); MOSTR (Sri Lanka); Swiss Funding Agencies (Switzerland); MST (Taipei); ThEPCenter, IPST, STAR, and NSTDA (Thailand); TUBITAK and TAEK (Turkey); NASU (Ukraine); STFC (United Kingdom); DOE and NSF (USA).

  \hyphenation{Rachada-pisek} Individuals have received support from the Marie-Curie program and the European Research Council and Horizon 2020 Grant, contract Nos.\ 675440, 724704, 752730, 758316, 765710, 824093, 884104, and COST Action CA16108 (European Union); the Leventis Foundation; the Alfred P.\ Sloan Foundation; the Alexander von Humboldt Foundation; the Belgian Federal Science Policy Office; the Fonds pour la Formation \`a la Recherche dans l'Industrie et dans l'Agriculture (FRIA-Belgium); the Agentschap voor Innovatie door Wetenschap en Technologie (IWT-Belgium); the F.R.S.-FNRS and FWO (Belgium) under the ``Excellence of Science -- EOS" -- be.h project n.\ 30820817; the Beijing Municipal Science \& Technology Commission, No. Z191100007219010; the Ministry of Education, Youth and Sports (MEYS) of the Czech Republic; the Deutsche Forschungsgemeinschaft (DFG), under Germany's Excellence Strategy -- EXC 2121 ``Quantum Universe" -- 390833306, and under project number 400140256 - GRK2497; the Lend\"ulet (``Momentum") Program and the J\'anos Bolyai Research Scholarship of the Hungarian Academy of Sciences, the New National Excellence Program \'UNKP, the NKFIA research grants 123842, 123959, 124845, 124850, 125105, 128713, 128786, and 129058 (Hungary); the Council of Science and Industrial Research, India; the Latvian Council of Science; the Ministry of Science and Higher Education and the National Science Center, contracts Opus 2014/15/B/ST2/03998 and 2015/19/B/ST2/02861 (Poland); the Funda\c{c}\~ao para a Ci\^encia e a Tecnologia, grant CEECIND/01334/2018 (Portugal); the National Priorities Research Program by Qatar National Research Fund; the Ministry of Science and Higher Education, projects no. 14.W03.31.0026 and no. FSWW-2020-0008, and the Russian Foundation for Basic Research, project No.19-42-703014 (Russia); MCIN/AEI/10.13039/501100011033, ERDF ``a way of making Europe", and the Programa Estatal de Fomento de la Investigaci{\'o}n Cient{\'i}fica y T{\'e}cnica de Excelencia Mar\'{\i}a de Maeztu, grant MDM-2017-0765 and Programa Severo Ochoa del Principado de Asturias (Spain); the Stavros Niarchos Foundation (Greece); the Rachadapisek Sompot Fund for Postdoctoral Fellowship, Chulalongkorn University and the Chulalongkorn Academic into Its 2nd Century Project Advancement Project (Thailand); the Kavli Foundation; the Nvidia Corporation; the SuperMicro Corporation; the Welch Foundation, contract C-1845; and the Weston Havens Foundation (USA).
\end{acknowledgments}

\bibliography{auto_generated} 

\cleardoublepage \appendix\section{The CMS Collaboration \label{app:collab}}\begin{sloppypar}\hyphenpenalty=5000\widowpenalty=500\clubpenalty=5000\cmsinstitute{Yerevan~Physics~Institute, Yerevan, Armenia}
A.~Tumasyan
\cmsinstitute{Institut~f\"{u}r~Hochenergiephysik, Vienna, Austria}
W.~Adam\cmsorcid{0000-0001-9099-4341}, J.W.~Andrejkovic, T.~Bergauer\cmsorcid{0000-0002-5786-0293}, S.~Chatterjee\cmsorcid{0000-0003-2660-0349}, K.~Damanakis, M.~Dragicevic\cmsorcid{0000-0003-1967-6783}, A.~Escalante~Del~Valle\cmsorcid{0000-0002-9702-6359}, R.~Fr\"{u}hwirth\cmsAuthorMark{1}, M.~Jeitler\cmsAuthorMark{1}\cmsorcid{0000-0002-5141-9560}, N.~Krammer, L.~Lechner\cmsorcid{0000-0002-3065-1141}, D.~Liko, I.~Mikulec, P.~Paulitsch, F.M.~Pitters, J.~Schieck\cmsAuthorMark{1}\cmsorcid{0000-0002-1058-8093}, R.~Sch\"{o}fbeck\cmsorcid{0000-0002-2332-8784}, D.~Schwarz, S.~Templ\cmsorcid{0000-0003-3137-5692}, W.~Waltenberger\cmsorcid{0000-0002-6215-7228}, C.-E.~Wulz\cmsAuthorMark{1}\cmsorcid{0000-0001-9226-5812}
% ...
\vskip\cmsinstskip
% ...
1:~Also at TU Wien, Wien, Austria\\
% ...
\end{sloppypar}
\end{document}